\documentclass[amssymb,amsmath,prd, a4paper, twocolumn,superscriptaddress,nofootinbib]{revtex4-1}

\usepackage{graphicx}
\usepackage{dcolumn}
\usepackage{bm}
\usepackage{graphicx}
\usepackage{array}
\usepackage{booktabs}
\usepackage {xcolor}
\usepackage{amssymb}
\usepackage{pifont}

\newcommand{\xmark}{\ding{55}}
\usepackage{multirow}
\usepackage{diagbox}
\usepackage{comment}
\usepackage{xcolor}
\usepackage{enumerate}
\usepackage{booktabs}
\usepackage[normalem]{ulem}
\usepackage[utf8]{inputenc}
\usepackage{hyperref}
\usepackage{enumerate}
\usepackage{cleveref}
\usepackage{amsmath}
\usepackage{commath}
\hypersetup{
    colorlinks = true,
    linkcolor = {blue},
    citecolor = {blue},
    urlcolor = {blue},
    linkbordercolor = {white},
}

\usepackage{siunitx}
\newcolumntype{C}[1]{>{\centering\arraybackslash}p{#1}}

\begin{document}

\def\nrtidalvthree{{\texttt{NRTidalv3}}}
\def\nrtidal{{\texttt{NRTidal}}}
\def\nrtidalvtwo{{\texttt{NRTidalv2}}}
\def\imrphenomdnrtidaltwo{{\texttt{IMRPhenomD\_NRTidalv2}}}
\def\imrphenomdnrtidalthree{{\texttt{IMRPhenomD\_NRTidalv3}}}
\def\imrphenomxasnrtidaltwo{{\texttt{IMRPhenomXAS\_NRTidalv2}}}
\def\imrphenomxasnrtidalthree{{\texttt{IMRPhenomXAS\_NRTidalv3}}}
\def\imrphenomxpnrtidalthree{{\texttt{IMRPhenomXP\_NRTidalv3}}}
\def\imrphenomxpnrtidaltwo{{\texttt{IMRPhenomXP\_NRTidalv2}}}
\def\imrphenompvtwonrtidaltwo{{\texttt{IMRPhenomPv2\_NRTidalv2}}}
\def\seobnrvfourt{{\texttt{SEOBNRv4T}}}
\def\lalsuite{{\texttt{LALSuite}}}
\def\seobnrvfiverom{{\texttt{SEOBNRv5\_ROM}}}
\def\seobnrvfourromnrtidalvtwo{{\texttt{SEOBNRv4\_ROM\_NRTidalv2}}}
\def\seobnrvfiveromnrtidalvtwo{{\texttt{SEOBNRv5\_ROM\_NRTidalv2}}}
\def\seobnrvfiveromnrtidalvthree{{\texttt{SEOBNRv5\_ROM\_NRTidalv3}}}
\def\teobresums{{\texttt{TEOBResumS}}}
\def\imrphenomxhm{{\texttt{IMRPhenomXHM}}}
\def\imrphenomxphm{{\texttt{IMRPhenomXPHM}}}
\def\imrphenomxphmST{{\texttt{IMRPhenomXPHM-SpinTaylor}}}
\def\seobnrvfivehmrom{{\texttt{SEOBNRv5HM\_ROM}}}
\def\imrphenomxhmnrtidalvthree{{\texttt{IMRPhenomXHM\_NRTidalv3}}}
\def\seobnrvfivehmromnrtidalvthree{{\texttt{SEOBNRv5HM\_ROM\_NRTidalv3}}}
\def\imrphenomxhmnrtidalvtwo{{\texttt{IMRPhenomXHM\_NRTidalv2}}}
\def\seobnrvfivehmromnrtidalvtwo{{\texttt{SEOBNRv5HM\_ROM\_NRTidalv2}}}
\def\imrphenomxas{{\texttt{IMRPhenomXAS}}}
\def\imrphenomxp{{\texttt{IMRPhenomXP}}}
\def\xhmnrtthree{{XHM\_NRT3}}
\def\xphmnrtthree{{XPHM\_NRT3}}
\def\eobhmnrtthree{{EOBv5HM\_NRT3}}
\def\xhmnrttwo{{XHM\_NRT2}}
\def\xasnrtthree{{XAS\_NRT3}}
\def\xpnrtthree{{XP\_NRT3}}
\def\eobnrtthree{{EOBv5\_NRT3}}
\def\xasnrttwo{{XAS\_NRT2}}
\def\equalas{{\texttt{M34q105as}}}
\def\unequalas{{\texttt{M35q227as}}}
\def\unequalprec{{\texttt{M35q227prec}}}
\def\seobnrvfivethm{{\texttt{SEOBNRv5THM}}}

\newcommand{\AEI}{\affiliation {Max Planck Institute for Gravitational Physics (Albert Einstein Institute), Am M\"uhlenberg 1, Potsdam 14476, Germany}}
\newcommand{\UP}{\affiliation {Institut f\"{u}r Physik und Astronomie, Universit\"{a}t Potsdam, Haus 28, Karl-Liebknecht-Str. 24/25, 14476, Potsdam, Germany}}
\newcommand{\UIB}{\affiliation{Departament de Física, Universitat de les Illes Balears, IAC3-IEEC, Crta. Valldemossa km 7.5, E--7122 Palma, Spain}}

\title{Leveraging \nrtidalvthree\ to develop gravitational waveform models\\ with higher-order modes for binary neutron star systems}

\author{Adrian Abac}
\email{adrian.abac@aei.mpg.de}
\AEI \UP

\author{Felip A. Ramis Vidal}
\UIB

\author{Marta Colleoni}
\UIB

\author{Anna Puecher}
\UP

\author{Alejandra Gonzalez}
\UIB

\author{Tim Dietrich}
\UP \AEI

\date{\today}

\begin{abstract}
Accurate and reliable gravitational waveform models are crucial in determining the properties of compact binary mergers. In particular, next-generation gravitational-wave detectors will require more accurate waveforms to avoid biases in the analysis. In this work, we extend the recent \nrtidalvthree\ model to account for higher-mode corrections in the tidal phase contributions for binary neutron star systems. The higher-mode, multipolar \nrtidalvthree\ model is then attached to several binary-black-hole baselines, such as the phenomenological \imrphenomxhm\ and \imrphenomxphm\ models, and  the effective-one-body-based model \seobnrvfivehmrom. We test the performance and validity of the newly developed models by comparing them with numerical-relativity simulations and other tidal models. Finally, we employ them in parameter estimation analyses on simulated signals from both comparable-mass and high-mass-ratio systems, as well as on the gravitational-wave event GW170817, for which we find consistent results with respect to previous analyses.
\end{abstract}

\maketitle

\section{\label{section: introduction}Introduction}
A new era of multi-messenger astronomy was ushered when the gravitational-wave (GW) event GW170817 was detected by the Advanced LIGO~\cite{LIGOScientific:2014pky} and Advanced Virgo~\cite{VIRGO:2014yos} detectors~\cite{LIGOScientific:2017vwq,LIGOScientific:2017ync,LIGOScientific:2017zic}. This was the first discovery of a GW signal emitted by the coalescence of a binary neutron star (BNS) system, and it was accompanied by a suite of electromagnetic counterparts ranging from radio waves to gamma rays~\cite{LIGOScientific:2017zic, LIGOScientific:2017ync}. Two years later, the LIGO-Virgo-KAGRA (LVK) Collaboration observed another event identified as a BNS merger, GW190425~\cite{LIGOScientific:2020aai}. Although the exact BNS merger rate is still unknown, in the future an increase in the number of BNS detection is foreseen, thanks to the improving sensitivity of current detectors and especially with the planned next-generation ones~\cite{KAGRA:2013rdx, Chan:2018csa, Lenon:2021zac, Branchesi:2023mws, Abac:2025saz}.

The detection of GW170817 already fostered numerous important scientific studies. Among these, it helped to constrain the equation of state (EOS) that governs matter in the interior of neutron stars (NSs)~\cite{LIGOScientific:2017vwq, Dietrich:2020efo}. Various theoretical and experimental approaches~\cite{Chin:1974sa, Serot:1997xg, Huth:2021bsp, Alford:2022bpp, Lattimer:2012nd, Lattimer:2021emm, Burgio:2021vgk, Zhu:2023ijx} try to describe the neutron-rich, supranuclear-dense matter inside NSs, and with the increasing detector sensitivity it is expected that the EOS will be further constrained with future observations, ruling out competing models~\cite{Shibata:2005xz, Lattimer:2012nd, Ozel:2016oaf, Koehn:2024set, Flanagan:2007ix, Hinderer:2009ca, LIGOScientific:2017vwq, LIGOScientific:2018hze, LIGOScientific:2020aai}. In GW signals, the EOS manifests itself through the NS's tidal deformability parameter $\Lambda$. This is a measure of the deformation of a NS in the presence of an external field, i.e., one that is generated by its companion, and leaves a specific signature in the emitted GWs~\cite{Flanagan:2007ix, Hinderer:2007mb}.

To estimate the source parameters of a BNS event, one usually employs Bayesian inference~\cite{Veitch:2014wba,Thrane:2018qnx} to compare the data gathered from the observations with predictions obtained by solving the Einstein Field Equations (EFEs). One avenue for these theoretical predictions is the use of numerical-relativity (NR) simulations. However, such simulations are known to come with a high computational cost and, therefore, typically solve only the last 10-20 orbits before merger~\cite{Hotokezaka:2015xka, Hotokezaka:2016bzh, Haas:2016cop, Kawaguchi:2018gvj, Kiuchi:2019kzt, Foucart:2018lhe, Dietrich:2018phi,Ujevic:2022qle, Gonzalez:2022mgo, Kuan:2024jnw, Hayashi:2024jwt, Kuan:2025bzu, Hayashi:2024jwt}.
Hence, it is necessary to construct waveform models that produce long, accurate waveforms while maintaining computational efficiency.

Analytical approaches to building waveform models include the Post-Newtonian (PN) formalism, which approximates the solution to the EFEs by treating the binary characteristic velocity $v$ as a perturbative parameter~\cite{Jaranowski:1997ky, Damour:2001bu, Blanchet:2001ax, Blanchet:2002av, Blanchet:2004ek}. In this framework, the corrections that describe tidal interactions enter at 5PN order~\cite{Vines:2011ud}, and are known up to 7.5PN~\cite{Damour:2012yf, Henry:2020ski, Narikawa:2023deu,Mandal:2024iug, Dones:2024odv}. However, PN approximations become increasingly inaccurate as $v$ approaches the speed of light $c$, i.e., when the separation between the two objects in the binary decreases. To remedy this, the PN information is often recast into an effective-one-body (EOB) formalism that treats the system as a point-particle embedded in an effective metric~\cite{Buonanno:1998gg, Buonanno:2000ef, Damour:2009zoi, Gamba:2022mgx, Gamba:2023mww, Bohe:2016gbl, Bernuzzi:2014owa,Nagar:2018zoe,Akcay:2018yyh,Gamba:2020ljo, Hinderer:2016eia, Steinhoff:2016rfi, Steinhoff:2021dsn, Haberland:2025luz, Albanesi:2025txj}. To date, there are two families of EOB models that describe BNS systems: \texttt{SEOBNR} and \texttt{TEOBResumS}, with their most recent models \texttt{SEOBNRv5THM}~\cite{Haberland:2025luz} and \texttt{TEOBResumS-Dalí}~\cite{Albanesi:2025txj}, respectively. EOB models involve solving a set of ordinary differential equations, which incurs large computational costs. However, their evaluation is accelerated with various methods, such as post-adiabatic approximations~\cite{Nagar:2018gnk,Mihaylov:2021bpf, Gamba:2020ljo}, reduced-order models (ROM)~\cite{Lackey:2016krb,Lackey:2018zvw, Purrer:2014fza, Purrer:2015tud, Pompili:2023tna}, and machine learning techniques~\cite{ Tissino:2022thn}, among others.

A different approach for constructing waveform approximants is employed by phenomenological models~\cite{Ajith:2009bn, Santamaria:2010yb, Kawaguchi:2018gvj, Dietrich:2017aum,Dietrich:2018uni,Dietrich:2019kaq,Abac:2023ujg, Colleoni:2025aoh, Williams:2024twp}, which take information from PN, EOB, and/or NR, and build a closed-form description of the tidal information that retains both accuracy and evaluation efficiency.
The calibration to NR data better captures the tidal phase contribution to the waveform in the late inspiral and close to merger. Examples of such (frequency-domain) models include the \texttt{KyotoTidal}~\cite{Kawaguchi:2018gvj} approximant, which is calibrated to NR simulations up to 1000 Hz, \texttt{PhenomGSF}~\cite{Williams:2024twp} model, calibrated to \texttt{TEOBResumS}~\cite{Gamba:2023mww} waveforms, and the \nrtidal\ series~\cite{Dietrich:2017aum,Dietrich:2018uni,Dietrich:2019kaq,Abac:2023ujg, Colleoni:2025aoh}. More recently, yet a different approach was proposed to calibrate phenomenological models directly to observational data~\cite{Abac:2025yml}.
The mentioned phenomenological models are modular in nature and can be augmented easily to any binary black hole (BBH) waveform baseline. 

In this paper, we focus on the \nrtidal\ waveform family, for which until now, extensions exist for \texttt{IMRPhenomD} \cite{Khan:2015jqa}, \texttt{IMRPhenomPv2} \cite{Hannam:2013oca, Khan:2018fmp}, \texttt{IMRPhenomXAS} \cite{Pratten:2020fqn}, \texttt{IMRPhenomXP} \cite{Pratten:2020ceb}, \texttt{SEOBNRv4\_ROM} \cite{Bohe:2016gbl}, and \texttt{SEOBNRv5\_ROM} \cite{Pompili:2023tna}. The \nrtidal\ models appended to the BBH baselines are publicly available in the LVK Algorithm Library (LAL)~\cite{lalsuite, swiglal}, and have been employed to analyze real GW data~\cite{LIGOScientific:2018hze, LIGOScientific:2017vwq, Dietrich:2019kaq, Abac:2023ujg, Colleoni:2024knd}. In particular, the most recent version, \nrtidalvthree~\cite{Abac:2023ujg}, improves on previous ones by using a larger set of NR data~\cite{Kiuchi:2017pte, Kawaguchi:2018gvj, Kiuchi:2019kzt, Dietrich:2018phi, Ujevic:2022qle, Gonzalez:2022mgo}, a mass-ratio-dependent fit between its parameters, and by introducing a frequency-dependent dynamical tidal deformability parameter $\Lambda\rightarrow \Lambda(f)$ to incorporate dynamical effects explicitly~\cite{Abac:2023ujg}. Additionally, it also retains the spin and amplitude corrections that were included in previous versions~\cite{Dietrich:2017aum, Dietrich:2018uni, Dietrich:2019kaq}.

However, the existing phenomenological BNS models only include tidal effects for the dominant $(2,|2|)$-mode. With increasing detector sensitivity, and in light of next-generation detectors, subdominant or higher-order modes (HM)~\cite{London:2017bcn, Khan:2015jqa,Cotesta:2020qhw} might play a more significant role, also for NSBH systems, which can be highly asymmetric in their mass ratios. For this reason, in this work, we introduce an update to \nrtidalvthree\ by incorporating HM effects, and attach this ``generalized" \nrtidalvthree\ to state-of-the-art HM BBH baselines, such as the aligned-spin models \imrphenomxhm~\cite{Garcia-Quiros:2020qpx} and \seobnrvfivehmrom~\cite{Pompili:2023tna, Khalil:2023kep, vandeMeent:2023ols, Ramos-Buades:2023ehm}, and the precessing model \imrphenomxphm~\cite{Pratten:2020ceb}, providing a more complete description of BNS systems that includes both dominant- and subdominant-mode effects. This is the first time that phenomenological tidal models are extended to include HM corrections, although HM extensions already exist for EOB BNS models, i.e., the aligned-spin \seobnrvfivethm~\cite{Haberland:2025luz} and the precessing and eccentric \texttt{TEOBResumS-Dalí}~\cite{Albanesi:2025txj}.

The paper is structured as follows. We discuss how the HM version of \nrtidalvthree\ is constructed in Sec.~\ref{section: Model Construction}. The implementation of the HM \nrtidalvthree\ model attached to BBH baselines, as well as tests on their validity and performance, is discussed in Sec.~\ref{section: Implementation and Performance}, with additional comparisons in Appendices~\ref{subsection: appA} and \ref{subsection: appA1}. We then perform parameter estimation (PE) using the higher-mode models for different systems in Sec.~\ref{section: Parameter estimation}, with an additional recovery in Appendix~\ref{subsection: appB}, and finally we conclude in Sec.~\ref{section: Conclusions and Outlook}, where we provide some outlook, including its use for the development of NSBH models.

Throughout this work, we employ geometric units $G = c = 1$ unless otherwise stated. We define the primary (heavier) mass in the BNS as $M_A$ and the secondary mass $M_B$, both in units of solar mass $M_{\odot}$. The tidal deformabilities and aligned-spin components are denoted as $(\Lambda_A,\,\, \chi_A)$ and $(\Lambda_B,\,\, \chi_B)$, respectively, and the mass ratio is defined as $q = M_A/M_B \ge 1.$ The total mass is $M = M_A + M_B$, from which we can construct the normalized masses $X_{A,B} = M_{A,B}/M$. For convenience, we refer to the `higher-order mode waveforms' as just the `HM waveforms', and we resort to shorthand names for all the approximants used throughout this study, which are summarized in Table~\ref{table: approximants_table}.

\section{\label{section: Model Construction} Model Construction}
In this section, we start with a brief overview of the \nrtidalvthree\ model~\cite{Abac:2023ujg} before discussing its extension to include HM corrections. 

\subsection{\label{subsection: NRTidalv3 Review} Review of the \nrtidalvthree\ model}

The frequency-domain (FD) $(2,|2|)$-mode GW strain can be written as
\begin{equation}
    h_{22}(f) = A_{22}(f)e^{-i\psi_{22}(f)},
\end{equation}
where $A_{22}(f)$ is the amplitude as a function of the GW frequency $f$ and the GW phase $\psi_{22}(f)$ can be approximated as
\begin{equation} 
    \psi_{22}(f) = \psi_{22}^0(f) + \psi_{22}^{\text{SO}}(f) + \psi_{22}^{\text{SS}}(f) +\psi_{22}^{\text{S}^3}(f) +  \psi_{22}^T(f) + ...,
\end{equation}
with $\psi_{22}^0$ denoting the point-mass contribution, $\psi_{22}^{\text{SO}}$ the spin-orbit coupling, $\psi_{22}^{\text{SS}}$ the spin-spin coupling, $\psi_{22}^{\text{S}^3}$ the spin-cubed term and $\psi_{22}^T$ the tidal phase~\cite{Flanagan:2007ix}. A similar expression and phase decomposition hold in the time domain (TD).

As discussed above, the tidal phase is known analytically up to 7.5PN order~\cite{Vines:2011ud, Damour:2012yf, Henry:2020ski, Narikawa:2023deu,Mandal:2024iug, Dones:2024odv}. However, the PN description of the tidal phase contribution cannot sufficiently capture the tidal effects in the late inspiral and up to merger. In \nrtidalvthree~\cite{Abac:2023ujg}, this is addressed by extending the 7.5PN tidal phase via calibration to NR simulations.

The dominant-mode tidal phase $\psi_T^{22}(f)$ is modeled by the \nrtidalvthree\ phase $\psi_T^{\rm NRT3}(x)$ as:
\begin{equation}\label{eq: Psi_TNRT3}
    \psi_T^{\text{NRT3}} = -\bar{\kappa}_A(\hat{\omega})\bar{ c}_{\text{Newt}}^Ax^{5/2} \bar{P}_{\text{NRT3}}^A(x) + [A\leftrightarrow B],
\end{equation}
where $\bar{ c}_{\text{Newt}}^A$ is the mass-ratio dependent leading-order Newtonian coefficient, $x = (\hat{\omega}/2)^{2/3}$ is the PN (frequency) parameter, $\hat{\omega} = M\omega = M(2\pi f)$ is the mass-weighted GW frequency, and $\bar{\kappa}_A(\hat{\omega}) = 3X_BX_A^4\Lambda_A(\hat{\omega})$ is the dynamical tidal parameter.  The dynamical tidal effects are encoded in the correction factor $\bar{k}_2^{\rm eff}(\hat\omega)$ which is multiplied to the adiabatic tidal deformability, i.e., $\Lambda_{A,B}(\hat{\omega}) = \Lambda_{A,B}\bar{k}_2^{\rm eff}(\hat\omega)$, with $\bar{k}_2^{\rm eff}(\hat\omega)\rightarrow 1$ at large separations. The Padè approximant $\bar{P}_{\rm NRT3}^A(x)$ is given by
\begin{equation}\label{eq: Padefreq}
\begin{split}
        &\bar{P}_{\text{NRT3}}^A(x)\\
        &= \frac{1 + \bar{n}_1^A x + \bar{n}_{3/2}^Ax^{3/2} + \bar{n}_2^Ax^2 + \bar{n}_{5/2}^Ax^{5/2} + \bar{n}_3^Ax^3}{1 + \bar{d}_1^Ax + \bar{d}_{3/2}^A x^{3/2}}.
\end{split}
\end{equation}
The following coefficients of Eq.~\eqref{eq: Padefreq} are fitted into the 7.5PN tidal phase\footnote{The fitting procedure includes Taylor-expanding $\bar{P}_{\text{NRT3}}^{A,B}(x)$ and matching the coefficients with the 7.5PN phase.}:
\begin{equation}
    \begin{split}
        \bar{n}_1^A &= \bar{c}_1^A + d_1^A,\\
        \bar{n}_{3/2}^A &= \frac{\bar{c}_1^A\bar{c}_{3/2}^A - \bar{c}_{5/2}^A - \bar{c}_{3/2}^A\bar{d}_1^A + \bar{n}_{5/2}^A}{\bar{c}_1^A},\\
        \bar{n}_2^A &= \bar{c}_2^A + \bar{c}_1^A \bar{d}_1^A\\
        \bar{d}_{3/2}^A &= -\frac{\bar{c}_{5/2}^A + \bar{c}_{3/2}^A\bar{d}_1^A - \bar{n}_{5/2}^A}{\bar{c}_1^A},
    \end{split}
\end{equation}
while the rest of the coefficients in Eq.~\eqref{eq: Padefreq} are given by 
\begin{equation}\label{eq: freeconstraint}
\begin{split}
    \bar{n}_i^{A,B}(\hat{\omega}) &= \bar{ a}_{i,0} + \bar{a}_{i,1} X_{A,B} \\
    &+ \bar{a}_{i,2} (\kappa_{A,B}+1)^{\bar{\alpha}}\\
    &+ \bar{a}_{i,3} X_{A,B}^{\bar{\beta}}, \quad \text{for}\quad  i \in [5/2, 3]\\
    \bar{d}_1^{A,B} &= \bar{d}_{1,0} + \bar{d}_{1,1} X_{A,B} + \bar{d}_{1,2} X_{A,B}^{\bar{\beta}},
\end{split}
\end{equation}
where $\bar{c}_{i,j}$ are the PN coefficients, while $\bar{\alpha}$, $\bar{\beta}$, $\bar{a}_{i,j}$ and $\bar{d}_{i,j}$ are calibrated to 55 non-spinning NR simulations~\cite{Kiuchi:2017pte, Kawaguchi:2018gvj, Kiuchi:2019kzt, Dietrich:2018phi, Ujevic:2022qle, Gonzalez:2022mgo}\footnote{We refer the reader to Ref.~\cite{Abac:2023ujg} for more details on the model construction as well as the values of the calibration parameters.}. 
This NR set is spread among different mass-ratios and various EOS. The model has been tested and can be employed reliably\footnote{No divergences or pathological behaviors have been observed for configurations within this parameter space.} for $M_{A,B} = [0.5, 3.0]M_{\odot}$, $\Lambda_{A,B} = [0, 25000]$, and $|\chi_{A,B}| \le 0.7$.

\subsection{\label{subsection: Inspiral} The HM tidal phase}
To write the HM phase, we first recall that the GW strain can be decomposed in terms of spherical harmonic modes $h_{\ell m} (t)$ multiplied by their respective spin-weighted spherical harmonic functions ${}_{-2}Y_{\ell m}(\theta, \phi)$~\cite{Thorne:1980ru}
\begin{equation}
    h(t) = \sum_{\ell \ge 2}\sum_{m = -\ell}^{\ell}{}_{-2}Y_{\ell m}(\theta, \phi)h_{\ell m}(t),
\end{equation}
where $\theta$ is the polar angle, and $\phi$ is the azimuthal angle.

In the inspiral and up to merger, it has been found that the phase follows a scaling relation of the form~\cite{Jaranowski:1997ky, Damour:2001bu, Blanchet:2001ax, Blanchet:2002av, Blanchet:2004ek, CalderonBustillo:2015lrg, London:2017bcn,Khan:2015jqa,Cotesta:2020qhw}
\begin{equation}
    \phi_{\ell m}(t) = |m|\phi_{\rm orb}(t) + \Delta \phi_{\ell m},
\end{equation}
where $\phi_{\rm orb} \approx \phi_{22}/2$ is the orbital phase and $\Delta\phi_{\ell m}$ is a constant phase shift. Moreover, Ref.~\cite{Ujevic:2022qle} (specifically, Fig.~4) confirmed that these scaling relations also hold for the tidal phase contributions $\phi_T^{\ell m}$. 

Assuming the stationary phase approximation (SPA) in transforming the waveform from TD to FD, and using the \nrtidalvthree\ model, we can write~\cite{CalderonBustillo:2015lrg, Garcia-Quiros:2020qpx} the phase of each mode as 
\begin{equation}\label{eq: highermodephase}
    \psi_{\ell m}(f) = \psi_{\ell m}^{\rm BBH} (f) + \frac{|m|}{2}\psi_{T, {\ell m}}^{\rm NRT3}\left(\frac{2f}{|m|} \right),
\end{equation}
in the building of a HM tidal model for BNS systems in the FD. 
Furthermore, the EOS-dependent spin-squared terms up to 3.5PN order and the leading-order spin-cubed terms entering at 3.5PN order that are present in \nrtidalvthree\ are also included and appropriately scaled, as in the above relations.
The HM effects are expected to be more significant for high-mass-ratio systems, and depend on the inclination angle $\iota$. In particular, for a ``face-on" (``edge-on") system, i.e., with $\iota = 0\,\, {\rm or}\,\, \iota = \pi$ ($\iota = \pi/2\,\, {\rm or}\,\, \iota = 3\pi/2$), the even (odd) $m$-modes are expected to have a significant contribution, while the odd (even) $m$-mode contributions are significantly suppressed~\cite{Thorne:1980ru, Garcia-Quiros:2020qpx, Pompili:2023tna}. 

\subsection{\label{subsection: Post-Merger} Tapering the waveform beyond merger}
Throughout this work, we only consider the description of the waveform up to merger. Due to the nature of the rational function $\bar{P}_{\rm NRT3}^{A,B}(x)$ used in \nrtidalvthree, for some combination of masses and tidal deformabilities, the function can diverge at frequencies beyond the merger frequency $f_{\rm mrg}$~\cite{Abac:2023ujg}\footnote{Further tests in Ref.~\cite{Abac:2023ujg} confirm that this asymptotic behavior, should it exist, always occurs at frequencies higher than $f_{\rm mrg}$.}. These divergences can propagate to the HMs due to the frequency scaling, as shown in Eq.~\eqref{eq: highermodephase}. 
Therefore, for the HM model extension, beyond merger we linearly extrapolate the phase, leading to
\begin{equation}
    \psi_T =
    \begin{cases}
        \psi_T(f), & f \leq f_{\mathrm{mrg}} \\[0.5ex]
        \psi_T(f_{\mathrm{s}}) + 
        \frac{\psi_T(f_{\mathrm{mrg}}) - \psi_T(f_{\mathrm{s}})}{f_{\mathrm{mrg}}-f_{\mathrm{s}}}
        (f - f_{\mathrm{s}}), & f > f_{\mathrm{mrg}}
    \end{cases}
\end{equation}
where $f_{\rm mrg}$ and $f_{\rm s} = 0.99f_{\rm mrg}$ are the reference points of the extrapolation, and where we use $\psi_T = \psi_T^{\rm NRT3}$ for brevity. As in Ref.~\cite{Abac:2023ujg}, we employ the updated merger frequency fit provided in Ref.~\cite{Gonzalez:2022mgo}.

Moreover, after the merger, we taper the waveform amplitude with a Planck taper window~\cite{McKechan:2010kp}
\begin{equation} \label{eq: PlanckTaper}
\sigma({f}) = \begin{cases}
    0,  \quad & {f} \le {f}_1\\
    \left[1+\exp\left(\frac{{f}_2 - {f}_1}{{f} - {f}_1} + \frac{{f}_2 - {f}_1}{{f} - {f}_2}\right)\right]^{-1}, \quad & {f}_1 \le {f} \le {f}_2\\
    1, \quad & {f} \ge {f}_2,
\end{cases}
\end{equation}
in the frequency interval $[f_1, f_2] = [f_{\rm mrg}, 1.2f_{\rm mrg}]$, i.e., right after merger.

\subsection{\label{subsection: Amplitude corrections} Tidal amplitude corrections}
The amplitudes of the HM waveforms in the FD do not follow the straightforward scaling relation as the phase in Eq.~\eqref{eq: highermodephase}; thus amplitude correction terms are needed. For the $(2,|2|)$-mode, both the \imrphenomxhm\ and \seobnrvfivehmrom\ baselines use the NR-calibrated amplitude corrections that were included in \nrtidalvtwo\ and the original $(2,|2|)$-mode implementation of \nrtidalvthree. For the HMs, we incorporate tidal amplitude corrections up to 7.5PN~\cite{Dones:2024odv} to the \imrphenomxhm\ baseline, but refrain from
doing so to the \seobnrvfivehmrom\ baseline given the difficulties and large computational cost involved in adapting these expressions to the implementation of the latter\footnote{Adding the 7.5PN tidal amplitude corrections to the \seobnrvfivehmrom\ baseline augmented with HM \nrtidalvthree\ increases the computational time of a single waveform evaluation by about 15\%. This issue does not arise in the case of the \texttt{IMRPhenomX(P)HM} baseline due to differences in waveform implementation.}.

\subsection{Inclusion of precession effects}
\label{subsection: Precession Effects}
For a more complete description of the gravitational waveforms, we also consider the possibility of systems with spins misaligned with the total angular momentum, which leads to precession in the orbital dynamics. To model such systems, tidal corrections are propagated to the co-precessing-frame modes by means of the twisting-up construction of the \imrphenomxphm\ baseline model~\cite{Pratten:2020ceb}.
This baseline model provides different spin-precession prescriptions, one of which is particularly interesting for BNS systems, as it allows for the inclusion of matter effects in the spin-induced quadrupole moments involved in the precession dynamics~\cite{Colleoni:2025aoh}\footnote{The default spin-precession prescription of \imrphenomxphm\ includes double-spin effects obtained via a multiple scale analysis (MSA) approximation of the precession-averaged spin couplings. The more accurate prescription based on the numerical integration of the orbit-averaged SpinTaylorT4 equations, which can be more suited for BNS systems, can be activated by setting \texttt{PhenomXPrecVersion:\,320} and \texttt{PhenomXPFinalSpinMod:\,2} in the \texttt{LALDict}.}.

\section{\label{section: Implementation and Performance} Implementation and performance}

\begin{table*}[t]
\caption{\label{table: approximants_table} Summary of BBH+\nrtidal\ approximants with HM corrections and the corresponding $(2,|2|)$-mode models implemented in \lalsuite. For each approximant, we state its short-hand alias to be used for the rest of this work, the BBH baseline, the version of \nrtidal\ attached, the total number of $(\ell, |m|)$ modes $n_{\rm Modes}$, the type of tidal amplitude correction used, precession implementation, and its median computational time using an Apple M1 Pro processor.}
\begin{ruledtabular}
\begin{tabular}{*{7}{c}}
    Alias & BBH baseline & $\psi_T$ & $n_{\rm Modes}$  & Tidal Amplitude Correction & Precession & $\Delta T \,\,[{\rm s}]$\\
    \hline\hline
    \multicolumn{7}{c}{HM Waveforms}\\
    \hline
    \xhmnrtthree & \imrphenomxhm & \nrtidalvthree & 5 & $(2,|2|)$: NR-calibrated; HMs: 7.5PN & \xmark & 0.0253 \\
    \xphmnrtthree & \imrphenomxphm & \nrtidalvthree & 5 & $(2,|2|)$: NR-calibrated; HMs: 7.5PN & \checkmark & 0.111\phantom{0} \\
    \xhmnrttwo & \imrphenomxhm & \nrtidalvtwo & 5 & $(2,|2|)$: NR-calibrated; HMs: 7.5PN & \xmark & 0.0250 \\
    \eobhmnrtthree & \seobnrvfivehmrom & \nrtidalvthree & 7 & $(2,|2|)$: NR-calibrated & \xmark & 0.0414 \\
    \hline
    \multicolumn{7}{c}{$(2,|2|)$-Mode Waveforms}\\
    \hline
    \xasnrtthree~\cite{Abac:2023ujg} & \imrphenomxas & \nrtidalvthree & 1 & $(2,|2|)$: NR-calibrated & \xmark & 0.00590 \\
    \xpnrtthree~\cite{Abac:2023ujg} & \imrphenomxp & \nrtidalvthree & 1 & $(2,|2|)$: NR-calibrated & \checkmark & 0.0189\phantom{0} \\
    \xasnrttwo~\cite{Colleoni:2025aoh} & \imrphenomxas & \nrtidalvtwo & 1 & $(2,|2|)$: NR-calibrated & \xmark & 0.00655 \\
    \eobnrtthree~\cite{Abac:2023ujg} & \seobnrvfiverom & \nrtidalvthree & 1 & $(2,|2|)$: NR-calibrated & \xmark & 0.00957 \\
\end{tabular}
\end{ruledtabular}
\end{table*}

We augment the HM \nrtidalvthree\ to the publicly available aligned-spin BBH waveform models \imrphenomxhm~\cite{Garcia-Quiros:2020qpx} and \seobnrvfivehmrom~\cite{Pompili:2023tna, Khalil:2023kep, vandeMeent:2023ols, Ramos-Buades:2023ehm}, and to the precessing model \imrphenomxphm~\cite{Pratten:2020ceb}. 
Table~\ref{table: approximants_table} provides the complete names of these new BNS waveform models, their features (i.e., the number of modes included, whether or not precession is incorporated, the type of amplitude corrections applied), the aliases that will be used from this point on, and, for completeness, their corresponding $(2,|2|)$-mode approximants. To test the consistency in the implementation of the models, we also augmented a HM version of \nrtidalvtwo\ to \imrphenomxhm. These models are implemented in the FD, but are also made available in the TD via inverse Fourier transformation routines built into \lalsuite~\cite{lalsuite}. For the $\texttt{IMRPhenomX(P)HM}$ baseline, there are $n_{\rm Modes}=5$ available modes $(\ell, |m|) = [(2,|2|),\allowbreak (2,|1|),\allowbreak (3,|2|),\allowbreak (3,|3|),\allowbreak (4,|4|)]$, while for $\seobnrvfivehmrom$, we have $n_{\rm Modes}=7$ given by $(\ell, |m|) = [(2,|2|),\allowbreak (2,|1|),\allowbreak (3,|2|),\allowbreak (3,|3|),\allowbreak (4,|3|),\allowbreak (4,|4|),\allowbreak (5,|5|)]$.

In the remainder of this section, we compare the runtimes of the different HM models with respect to their $(2,|2|)$-mode counterparts, and we check their validity by comparing them with NR simulations and other tidal approximants.

\subsection{\label{subsection: Timings} Timings}
\begin{figure}
\centering
\includegraphics[width=\linewidth]{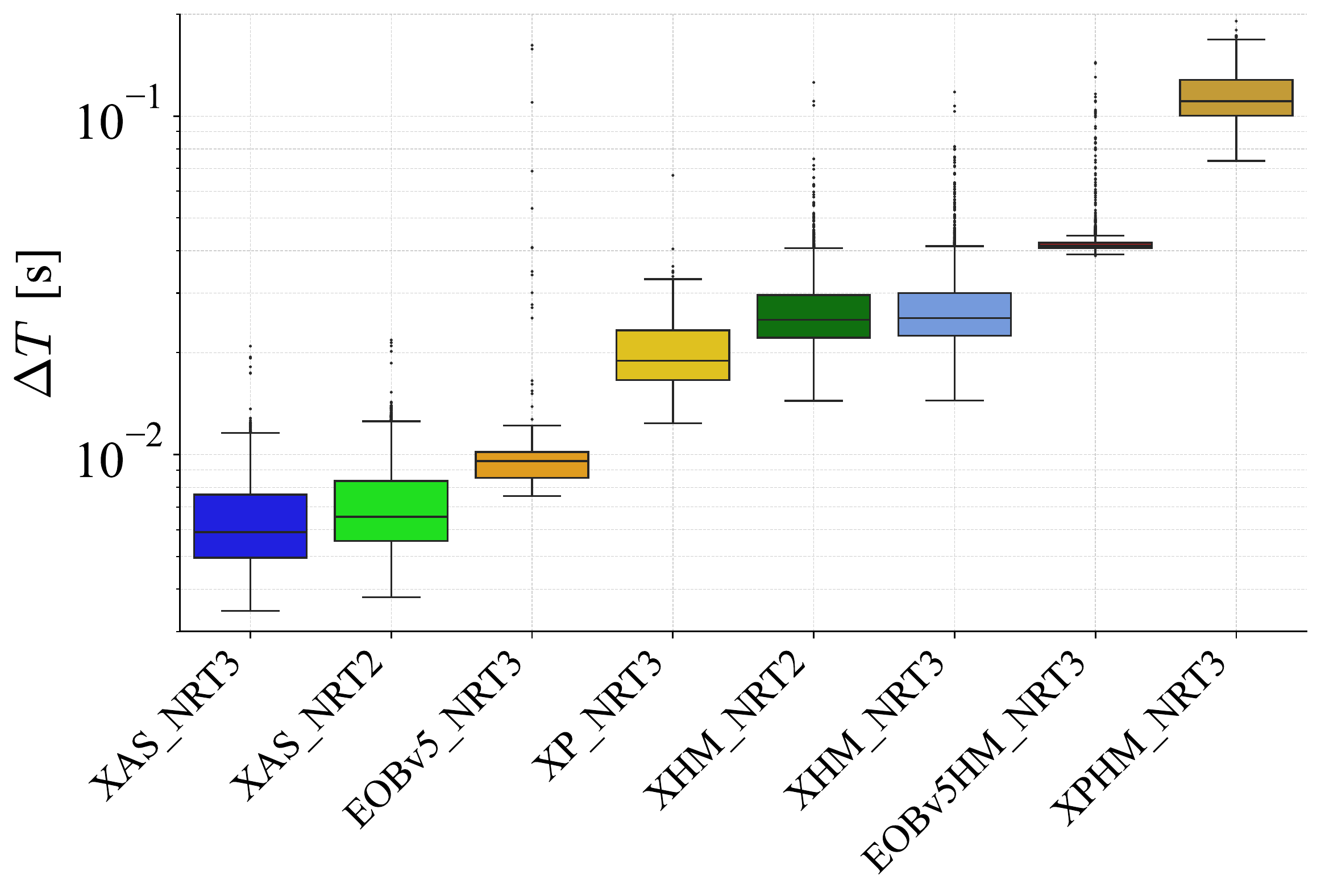}
\caption{Runtime boxplots of the different HM tidal models and their respective $(2,|2|)$-mode counterparts for the aligned spin and precessing models. Horizontal lines between the boxes indicate the median runtimes, whose values can be found in Table~\ref{table: approximants_table}. For each model, the box contains the middle 50\% of the distribution (between the first quartile $Q_1$ or the 25th percentile and the third quartile $Q_3$ or the 75th percentile), while the height of the box itself is the inter-quartile range $\Delta Q = Q_3 - Q_1$. The horizontal line in the middle of the box is the median of the distribution, while the lower and upper whiskers that extend from the box denote $Q_1 - 1.5\Delta Q$ and $Q_3 + 1.5\Delta Q$, respectively. Values higher or lower than the whiskers are denoted with individual points.}
\label{fig: timings}
\end{figure}
We compare the waveform evaluation times of \xhmnrtthree, \xphmnrtthree, \xhmnrttwo, and \eobhmnrtthree, with their respective $(2,|2|)$ mode versions \xasnrtthree, \xpnrtthree, \xasnrttwo, and \eobnrtthree. For all models, we generate 5000 random FD waveforms that include all their available modes and cover the parameter space $M_{A,B} = [0.5, 3.0]M_{\odot}$, $\Lambda_{A,B} = [0, 25000]$, and $|\chi_{A,B}| \le 0.7$, all with minimum frequency $f_{\rm min} = 20 \, \rm Hz$ and a frequency step $df = 0.125\, \rm Hz$. For the precessing models \xphmnrtthree\ and \xpnrtthree, we include spin precession parameters such that the component spin magnitudes still follow $|\chi_{A,B}| \le 0.7$. The median runtimes $\Delta T$ are reported in Table~\ref{table: approximants_table}, and their distributions are shown in Fig.~\ref{fig: timings} as box plots.

We observe from Table~\ref{table: approximants_table} and Fig.~\ref{fig: timings} that \xhmnrtthree\ has runtimes comparable to \xhmnrttwo\ despite \nrtidalvthree\ including more physics and having more complicated expression and implementation relative to \nrtidalvtwo. \eobhmnrtthree\ is $\sim 1.7$~times slower than the XHM models, but this is caused by the difference in speeds between their corresponding $(2,|2|)$-mode versions, attributed to the different implementation of the \imrphenomxhm\ and \seobnrvfivehmrom\ BBH baselines. Furthermore, comparing the performance of the HM and $(2,|2|)$-mode approximants, we find relative speed factors of 
\begin{equation}
\begin{split}
    \Delta T^{\rm \xhmnrtthree}/\Delta T^{\rm \xasnrtthree} &\approx 4.3\\ \Delta T^{\rm \xphmnrtthree}/\Delta T^{\rm \xpnrtthree} &\approx 5.9\\ \Delta T^{\rm \xhmnrttwo}/\Delta T^{\rm \xasnrttwo} &\approx 3.8\\ \Delta T^{\rm \eobhmnrtthree}/\Delta T^{\rm \eobnrtthree} &\approx 4.3. 
\end{split}
\end{equation}
For \xhmnrtthree\ and \eobhmnrtthree, we note that, despite including 5 and 7 modes respectively, the overhead relative to their non-HM counterparts remains small.

\begin{table*}
\caption{\label{table: bns_td_configs} BNS configurations of the NR simulations~\cite{Ujevic:2022qle, Dietrich:2017aum} used for the TD dephasing comparisons. For each configuration, we state the EOS that was used, individual masses $M_A$ and $M_B$, individual tidal deformabililities $\Lambda_A$ and $\Lambda_B$ as well as aligned-spin components $\chi_A$ and $\chi_B$. We also include $n_{\rm Modes}$ used from each NR waveform for the TD comparisons $n_{\rm Modes} = 5$ employs $(\ell, m) \in [(2,2), (2,1), (3,2), (3,3), (4,4)]$, while $n_{\rm Modes} = 5$ employs $(\ell, m) \in [(2,2), (3,2), (4,4)]$. Richardson-extrapolated waveforms of BAM:0131 and BAM:0130 were used in the TD comparisons, while the rest utilize the highest-resolution waveforms. The comparisons for the last four systems in the table are found in Appendix~\ref{subsection: appA}.}
\begin{ruledtabular}
    \begin{tabular}{l||c S[table-format=1.2] S[table-format=1.2] S[table-format=4.0] S[table-format=4.0] S[table-format=+1.2] S[table-format=+1.2] S[table-format=+1.2]}
        Name & EOS & $M_A$ [$M_{\odot}$] & $M_B$ [$M_{\odot}$] & $\Lambda_A$ & $\Lambda_B$ & $\chi_A$ & $\chi_B$ & $n_{\rm Modes}$ \\
        \hline
        BAM:0131                         & SLy  & 1.72 & 0.98 & 66   & 2557 & 0 & 0 & 5 \\
        BAM:0130                         & SLy  & 1.80 & 0.90 & 43   & 4088 & 0 & 0 & 5 \\
        BAM:0137                         & SLy  & 1.50 & 1.20 & 191  & 812  & 0 & 0 & 5 \\
        BAM:0136                         & SLy  & 1.62 & 1.08 & 108  & 1503 & 0 & 0 & 5 \\
        BAM:0062 & MS1b & 1.35 & 1.35 & 1531 & 1531 & -0.10 & -0.10 & 3 \\
        BAM:0101 & SLy & 1.35 & 1.35 & 390 & 390 & 0.05 & 0.05 & 3
    \end{tabular}
\end{ruledtabular}
\end{table*}
\subsection{\label{subsection: Time-Domain Comparisons} Time-domain comparisons}
\begin{figure*}
 \centering
\includegraphics[width=0.49\linewidth]{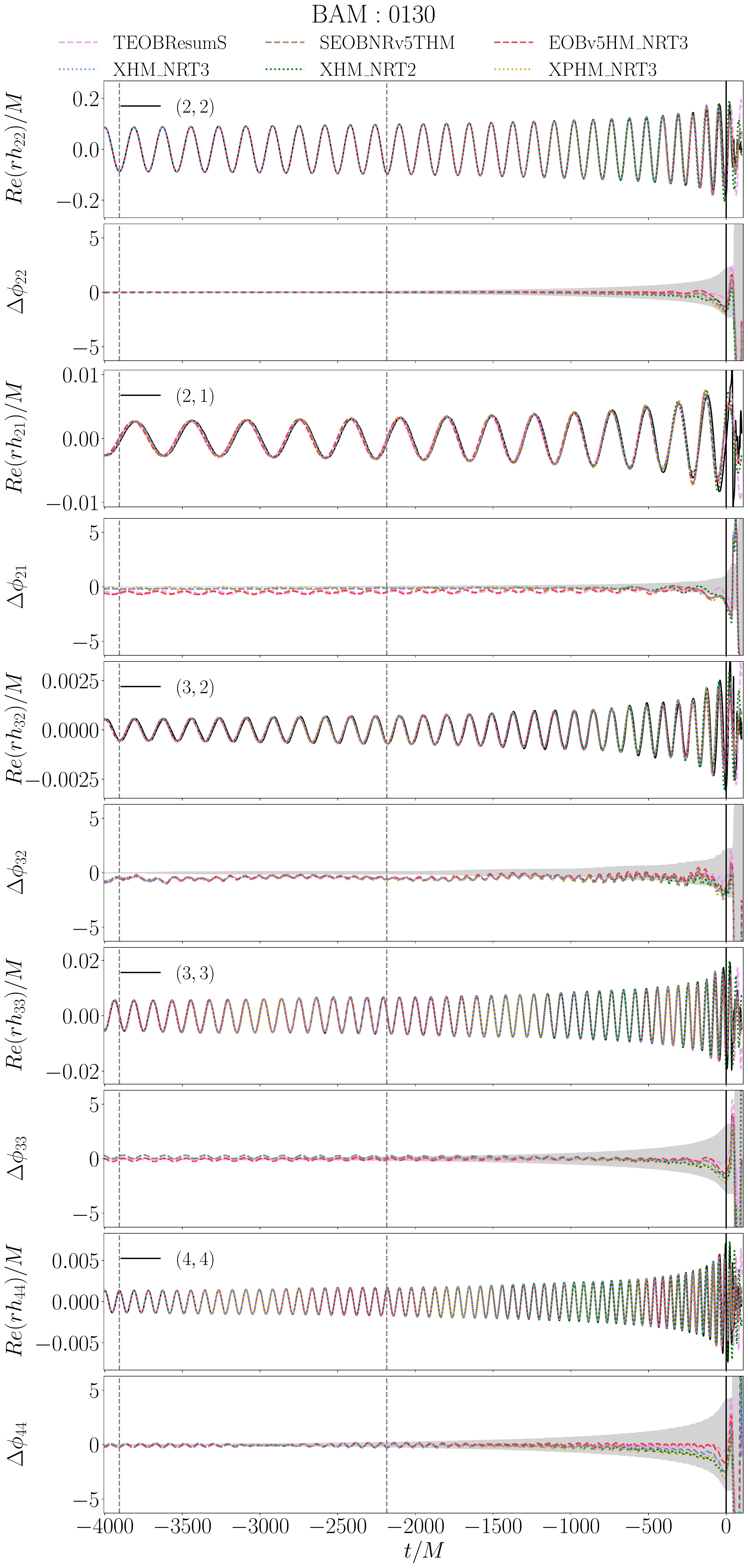}\hfill
\includegraphics[width=0.49\linewidth]{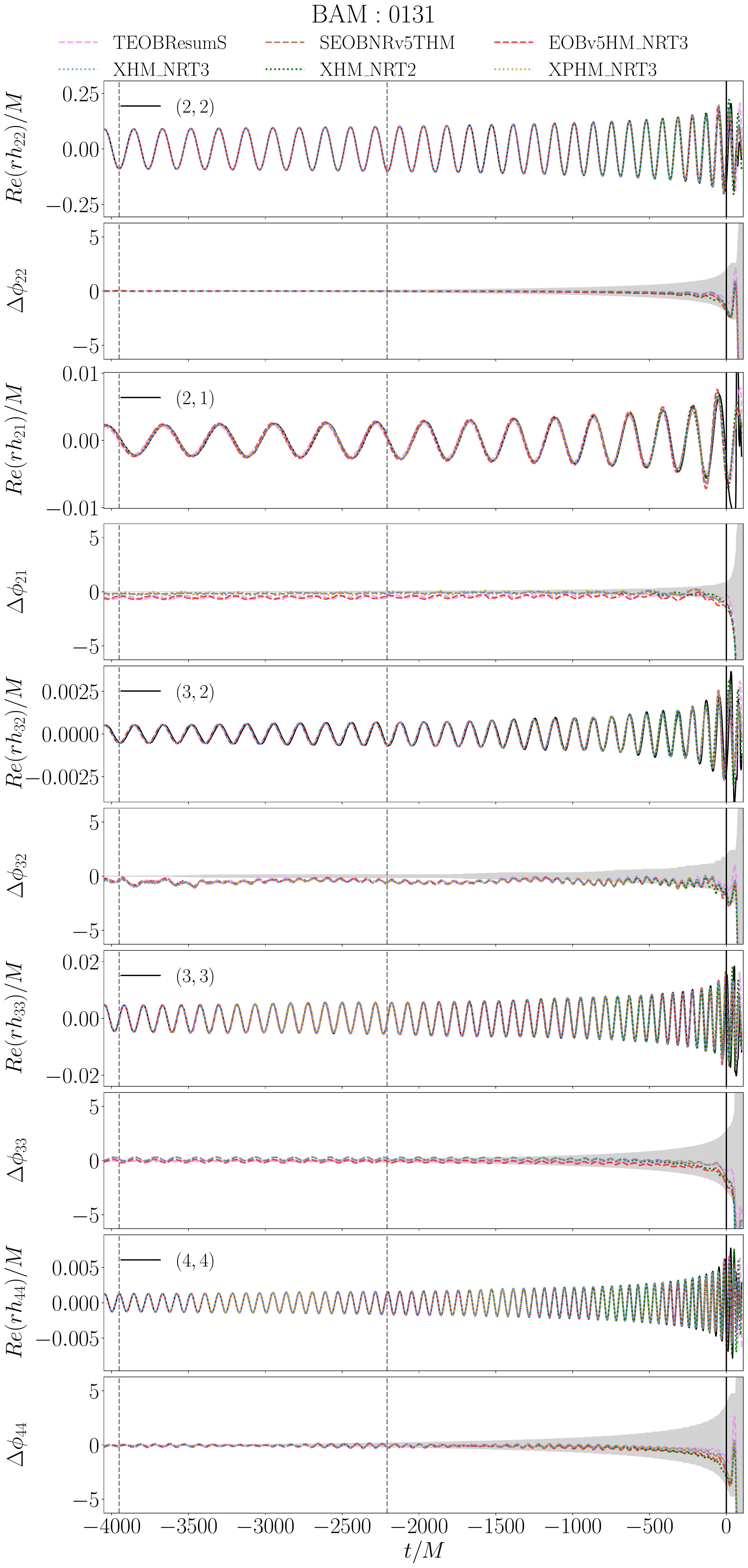}
\caption{Time-domain dephasing comparisons for the BAM:0130 and BAM:0131 waveforms. For each NR waveform, the upper panel shows the real part of the gravitational wave strain as a function of the retarded time, while the bottom panel shows the phase difference between the waveform model and the NR waveform. The gray band in the bottom panel represents the phase difference between the Richardson-extrapolated phase and the highest-resolution phase of the simulation. For each comparison, we denote the alignment windows by the dashed gray lines, and merger by the solid black line at $t/M = 0$.}
\label{fig: timedomaincomparisons1}
\end{figure*}

In the TD, we investigate the dephasing of the HM BNS models with respect to NR simulations that contain significant HM contributions. The simulations used for the comparison are listed in Table~\ref{table: bns_td_configs}. 
In particular, we consider four NR simulations with unequal masses (BAM:0130, BAM:0131, BAM:0136 and BAM:0137)~\cite{Ujevic:2022qle}, and two simulations of equal-mass systems (BAM:0062 and BAM:0101)~\cite{Dietrich:2017aum}; in the latter, the odd $m$ modes are suppressed due to the mass symmetry, but the even $m$-mode contributions are significant. In this section, we show the results for the comparison with unequal-mass simulations BAM:0130 and BAM:0131.
The comparative results for the other systems can be found in Appendix~\ref{subsection: appA}.
We also include in the comparisons the aligned-spin EOB models \texttt{TEOBResumS-Dalí}~\cite{Albanesi:2025txj} and \texttt{SEOBNRv5THM}~\cite{Haberland:2025luz}. For consistency, we restrict the comparison to the modes present in all models, i.e.,  $(\ell, m) \in [(2,2), (2,1), (3,2), (3,3), (4,4)]$. For the equal-mass systems, we further restrict ourselves to the even $m$-modes, i.e., $(\ell, m) \in [(2,2), (3,2), (4,4)]$.

To compare the different models with the NR simulations, we first need to align the waveforms over all their common modes by solving for the time and phase shifts $\delta t$ and $\delta \phi$, respectively, which minimize the following integral~\cite{Hotokezaka:2015xka, Hotokezaka:2016bzh}:
\begin{equation} \label{eq: alignment}
    \mathcal{I}(\delta t, \delta \phi) = \sum_{\ell m}\int_{t_1}^{t_2}{\!dt|\phi_{\ell m}^{\rm NR}(t) - \phi_{\ell m}^{\rm model}(t + \delta t) + m\delta \phi|},
\end{equation}
where $\phi_{\ell m}^{\rm NR}$ and $\phi_{\ell m}^{\rm model}$ are the phases of the NR and model waveforms, respectively, $\ell$ 
and $m$ run over the 5 or 3 modes listed above, and the integral runs over an arbitrary time window $[t_1, t_2]$, typically chosen to lie in the early part of the NR simulation. In this way, the phase of each mode is then rescaled by $e^{i m \delta \phi}$. The models are generated with a sampling rate $f_{s} = 2^{14} \,\rm{Hz} = 16384 \,\rm{Hz}$ and minimum frequency $f_{\rm min} = 40 \, \rm Hz$. 

Figure~\ref{fig: timedomaincomparisons1} shows the results of the mode-by-mode comparisons with BAM:0130 (left panel) and BAM:0131 (right panel). Due to their 2nd-order convergence, in the original calibration of \nrtidalvthree, Richardson-extrapolated waveforms of the $(2,|2|)$-modes of these waveforms were used~\cite{Abac:2023ujg}\footnote{See Ref.~\cite{Abac:2023ujg} for information of all the waveforms used in the calibration for the \nrtidalvthree\ tidal phase.}.
We find that this convergence can be extended into the HMs, so for the comparison with these waveforms we employ a Richardson-extrapolated versions of all the modes. 
The gray bands denote the uncertainty in the NR simulations, estimated as the phase difference between the Richardson-extrapolated phase and highest resolution phase.
From Fig.~\ref{fig: timedomaincomparisons1}, we observe that the developed HM models generally agree well, i.e., within the NR uncertainty, with the NR waveforms, particularly in the late inspiral. For some of the modes considered (c.f., the $(2,|1|)$ and $(3,|2|)$-modes), we observe a slight disagreement between the NR waveforms and the approximants, which could be due to the noise of the NR simulation, the underestimated error between resolutions, or the global minimization that was carried out in the alignment procedure. Performing an individual, mode-by-mode alignment (i.e., independently aligning each mode and calculating individual values for $\delta t$ and $\delta \phi$ for each mode, for every model) results in a very good agreement between the models and NR waveforms also in the early inspiral. However, this approach is not applicable and neglects the existing symmetry between the individual modes~\cite{Garcia-Quiros:2020qpx, Pratten:2020ceb}.

In addition to the aligned-spin NR waveforms listed in Table~\ref{table: bns_td_configs}, we also compare the \xphmnrtthree\ and \xpnrtthree\ models with precessing NR waveforms. The results are shown in Appendix~\ref{subsection: appA1}. In general, the \xphmnrtthree\ model agrees well with the NR waveforms while \xpnrtthree\ is in good agreement with the $(2,|2|)$-modes of these waveforms.

\subsection{\label{subsection: Mismatch Comparisons} Frequency-domain comparisons}
\begin{figure}
\centering
\includegraphics[width=\linewidth]{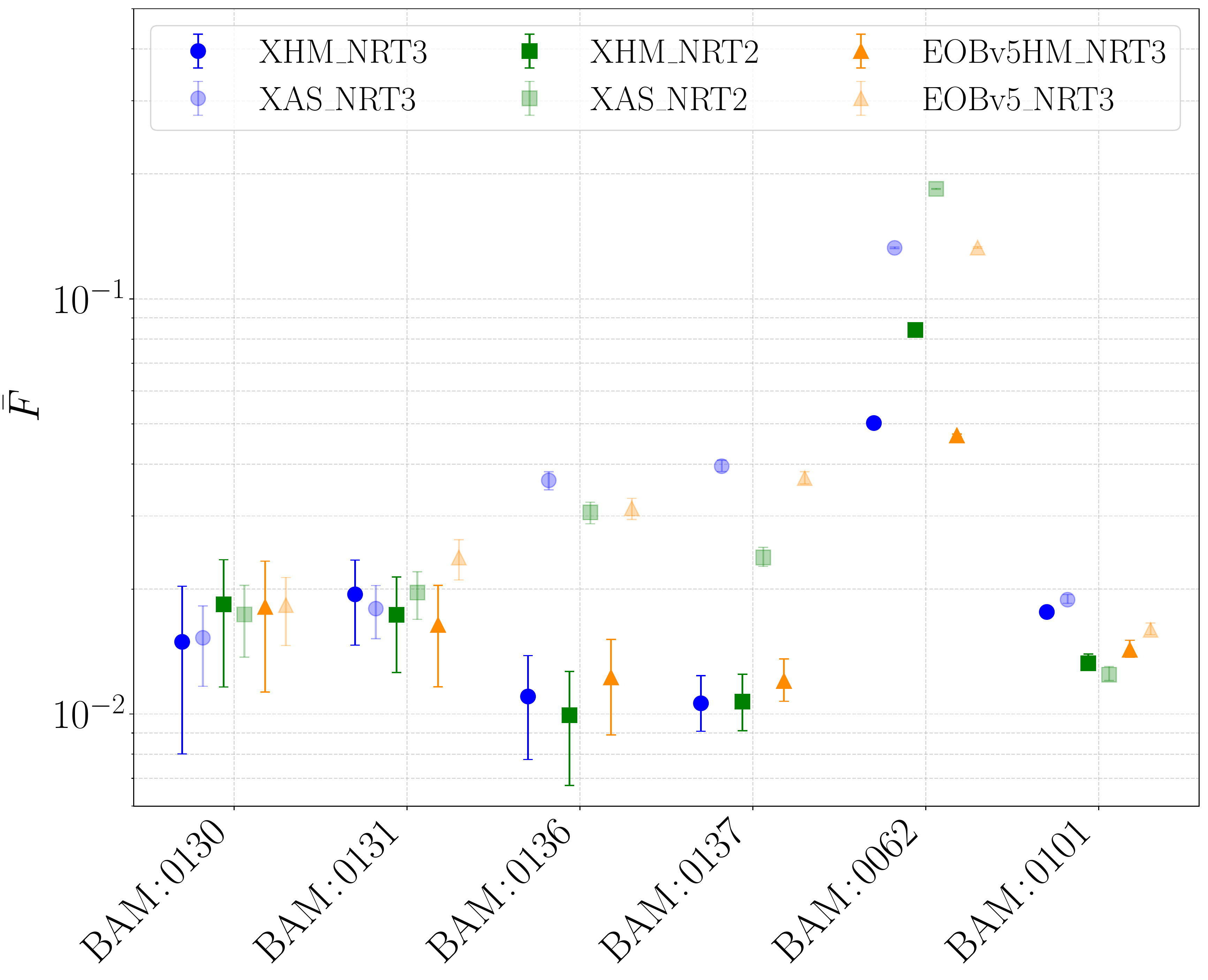}
\caption{Mismatches between the waveform models and the NR waveforms used in the time-domain comparisons (see Table~\ref{table: bns_td_configs}). For each model, the marker denotes the mean mismatch over 18 combinations of the inclination $\iota$, reference phase $\phi_0$, and polarization angle $\psi_p$. The upper and lower whiskers denote the maximum and minimum mismatches, respectively. For comparison, for every HM model we also show its $(2,|2|)$-mode counterpart, with the same marker and color but increased transparency in the plot. A smaller variability in the mismatches is observed for the $(2,|2|)$-mode models, and for the equal-mass simulations BAM:0062 and BAM:0101 due to the suppressed HM content.}
\label{fig: NRmismatches}
\end{figure}
\begin{figure}
\includegraphics[width=\linewidth]{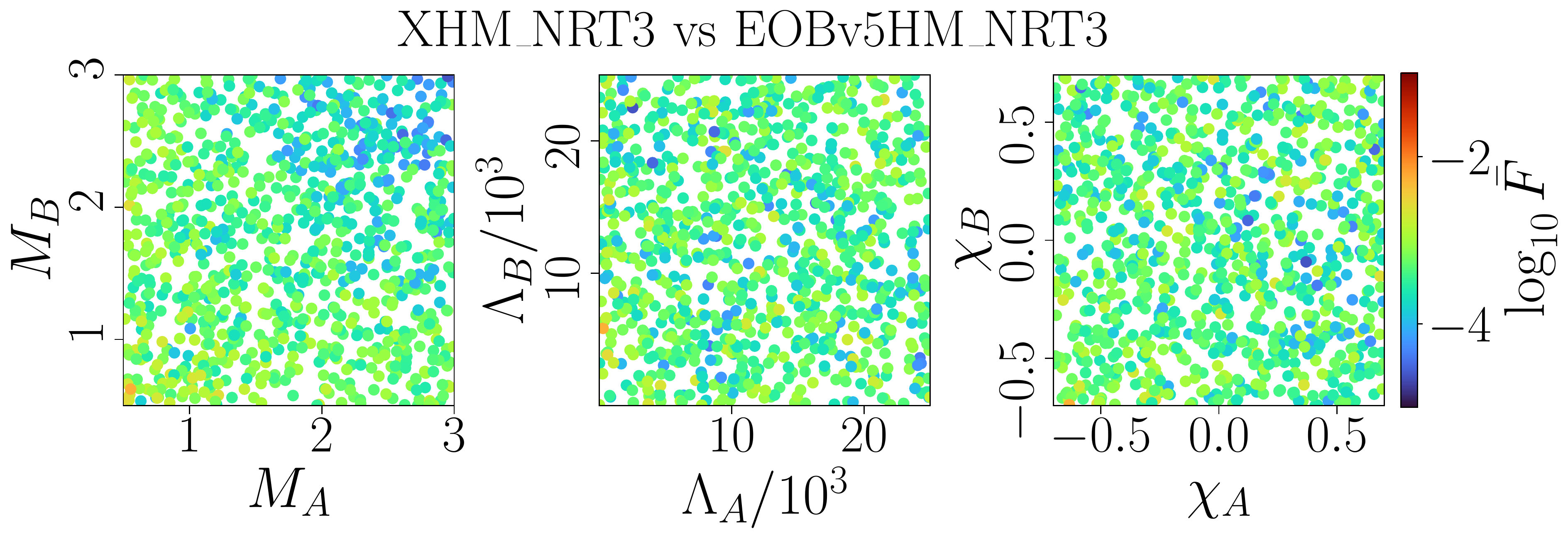}
\includegraphics[width=\linewidth]{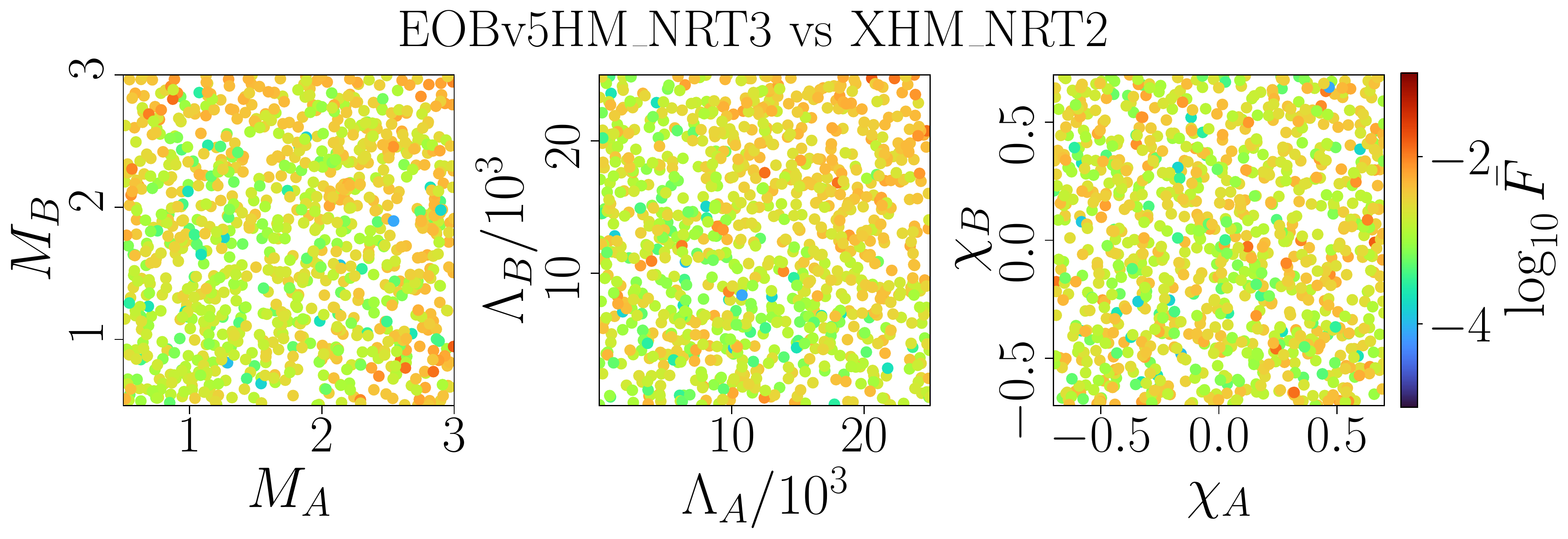}
\includegraphics[width=\linewidth]{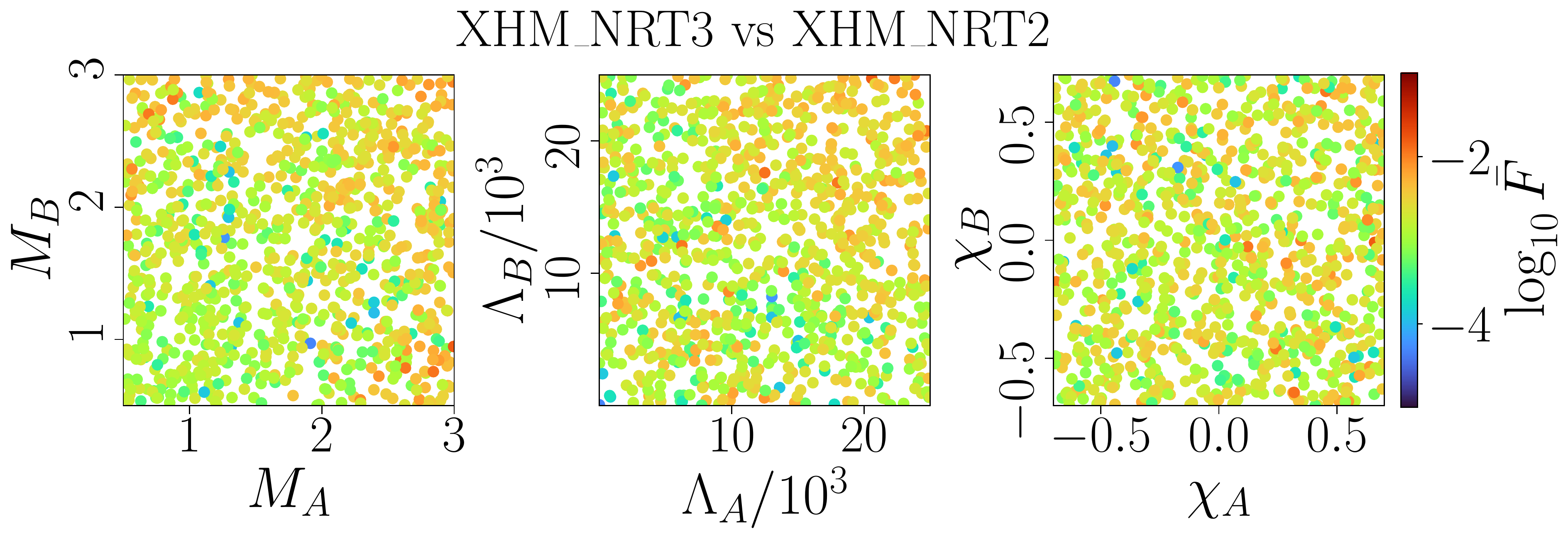}
\includegraphics[width=\linewidth]{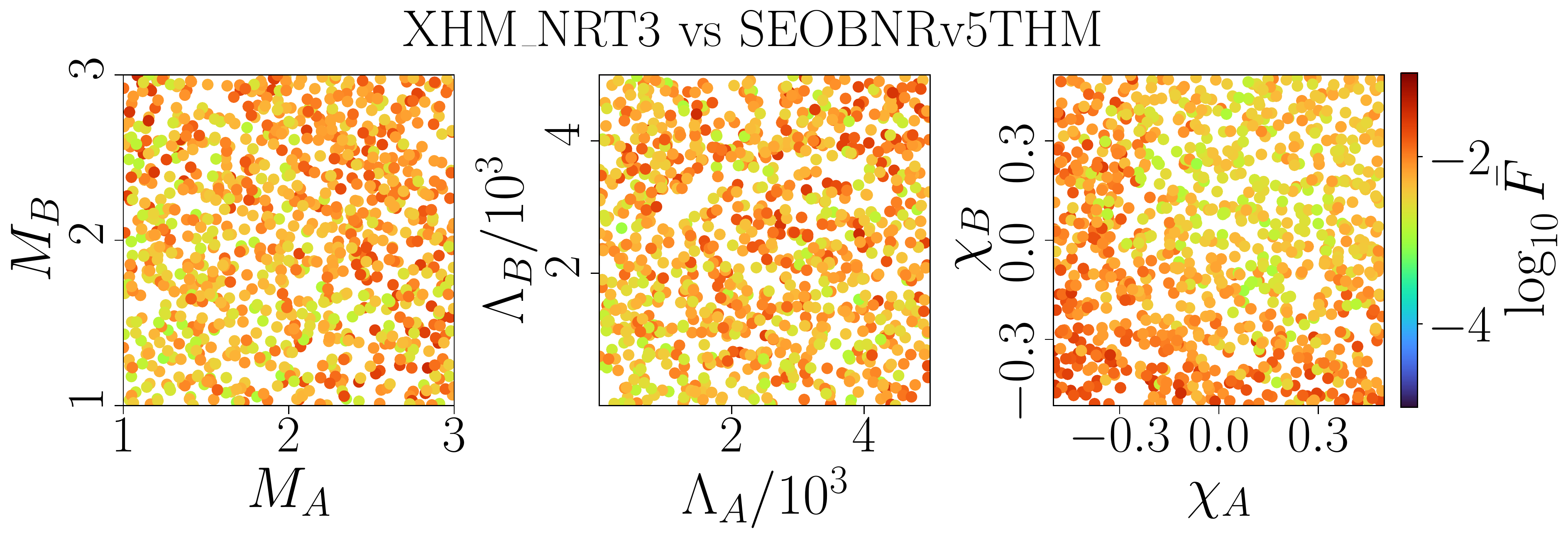}
\includegraphics[width=\linewidth]{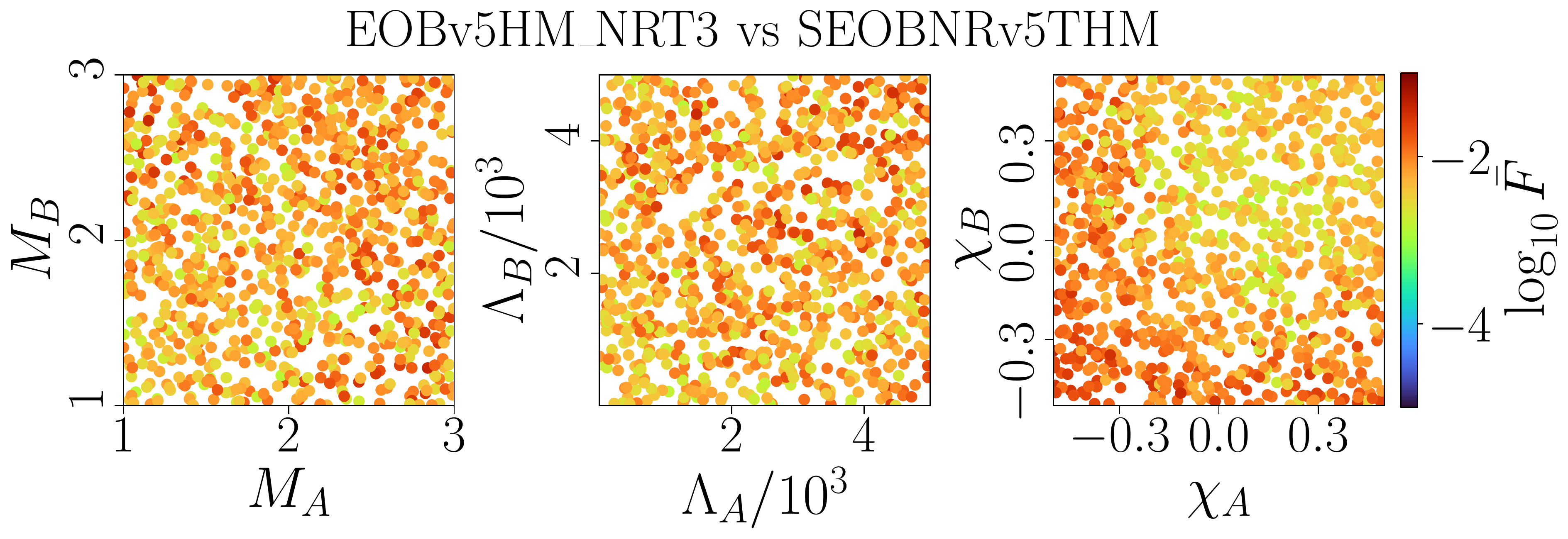}
\caption{Mismatch comparisons of between the different HM tidal waveform models. Each subfigure contains three two-dimensional, density scatter plots of the masses $M_{A,B}$, tidal deformabilities $\Lambda_{A,B}$, and aligned spin components $\chi_{A,B}$, where the log-mismatch $\log_{10}\bar{F}$ is represented by the color bar, with limits set to $\log_{10}\bar{F} \in [-5, -1]$. We note the smaller axis limits for the comparison with \texttt{SEOBNRv5THM}, due to the narrower parameter space used here.}
\label{fig: mm_allmodes}
\end{figure}

\begin{figure}
\centering
\includegraphics[width=\linewidth]{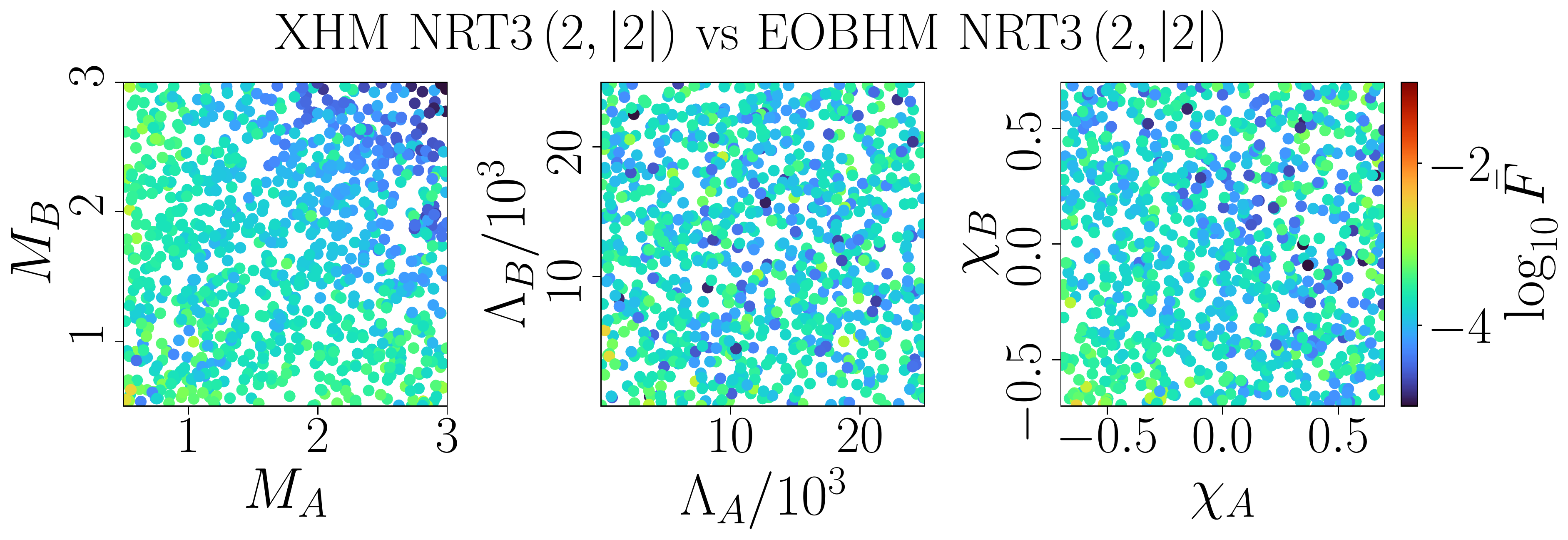}
\includegraphics[width=\linewidth]{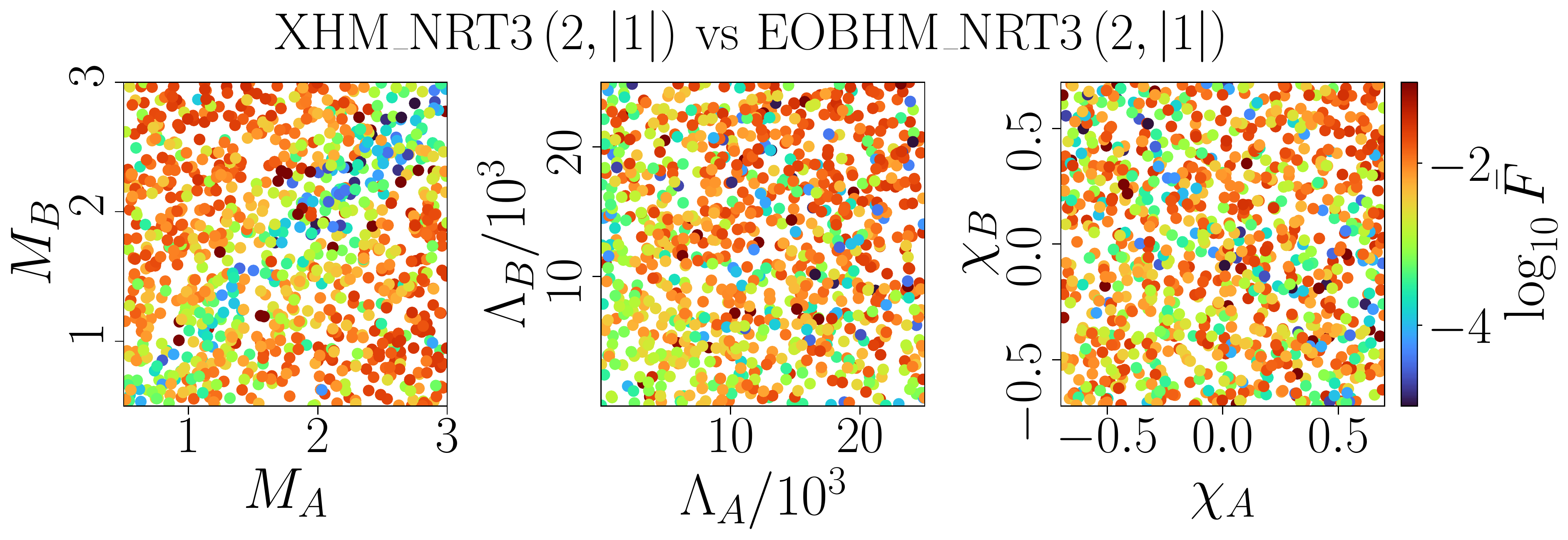}
\includegraphics[width=\linewidth]{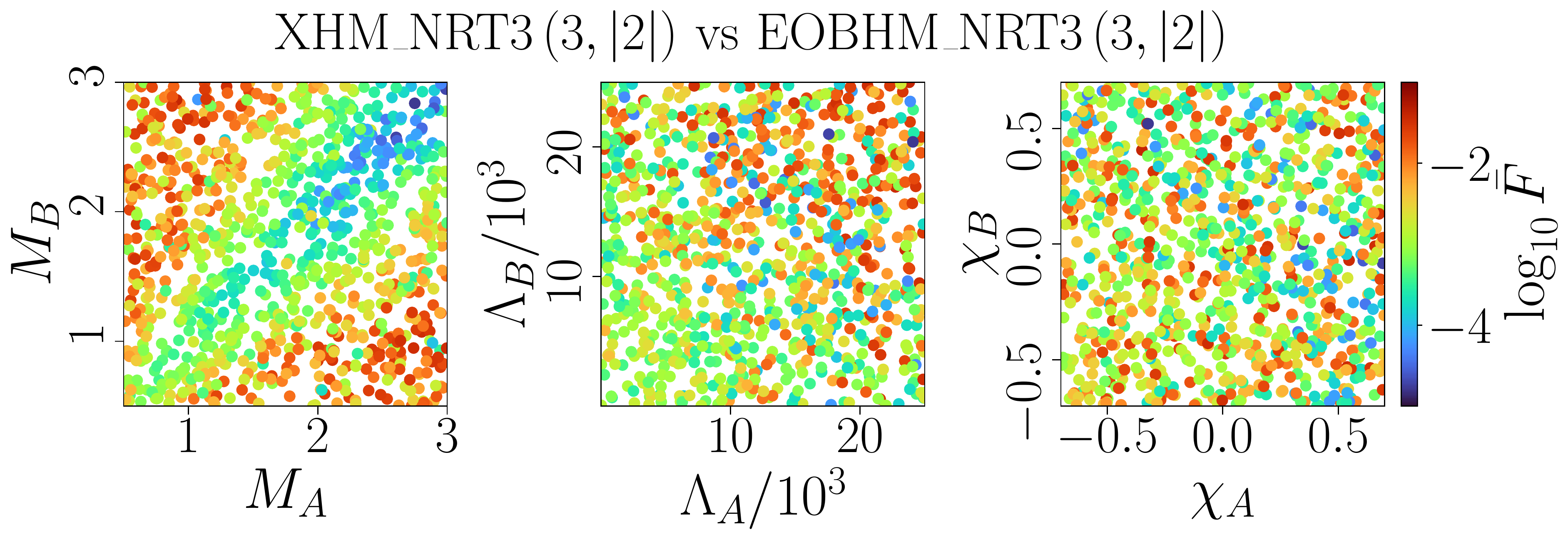}
\includegraphics[width=\linewidth]{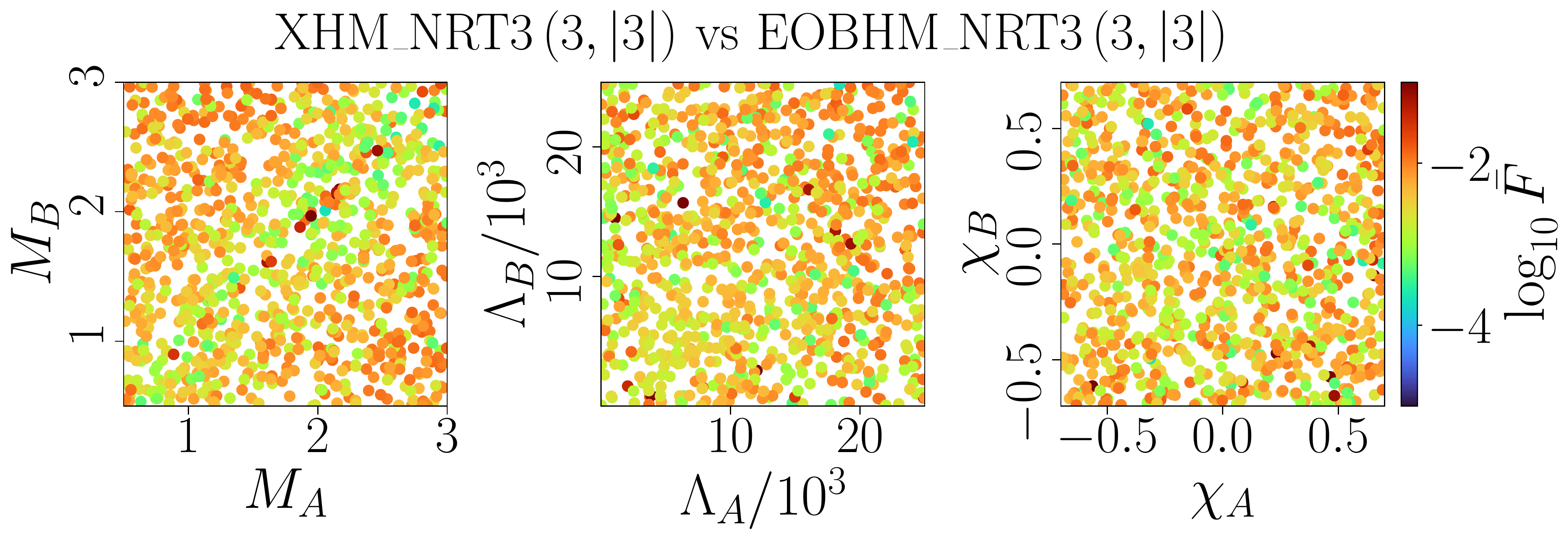}
\includegraphics[width=\linewidth]{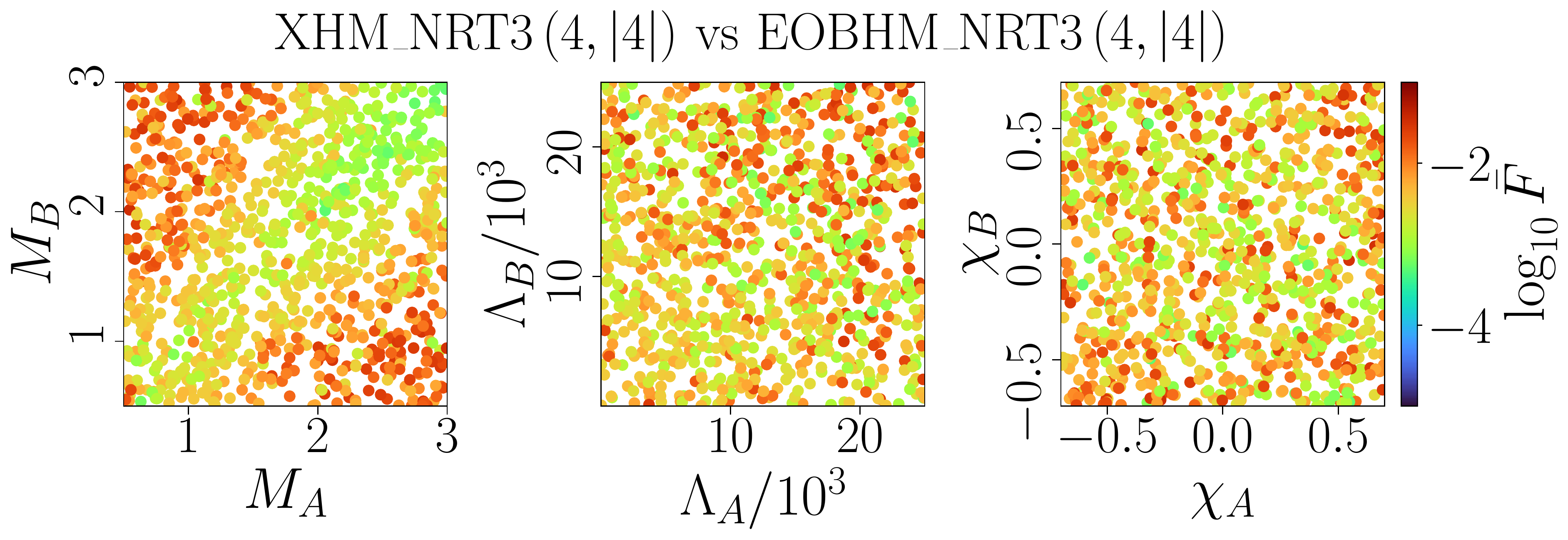}
\caption{Mismatch comparisons per mode between \xhmnrtthree\ and \eobhmnrtthree. Each common mode is represented by a subfigure that contains three two-dimensional, density scatter plots of the masses $M_{A,B}$, tidal deformabilities $\Lambda_{A,B}$, and aligned spin components $\chi_{A,B}$, where the log-mismatch $\log_{10}\bar{F}$ is represented by the color bar, with limits set to $\log_{10}\bar{F} \in [-5, -1]$.}
\label{fig: mode_by_mode_mm}
\end{figure}

\begin{figure}
\includegraphics[width=\linewidth]{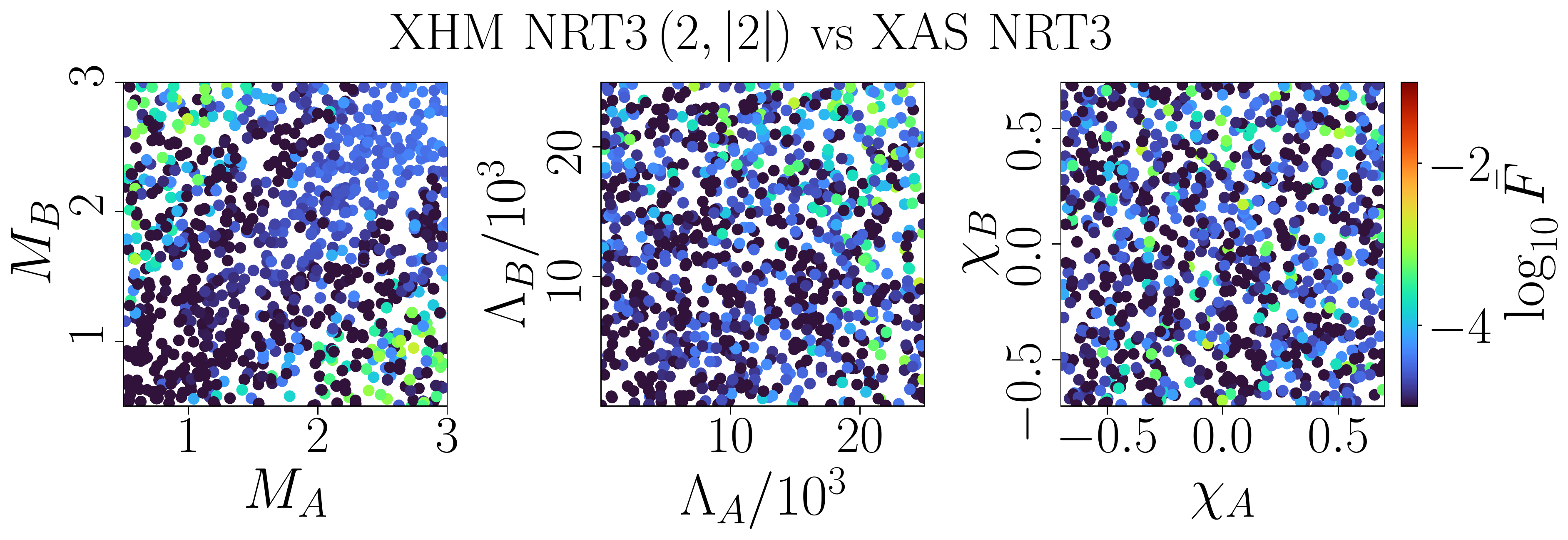}  \includegraphics[width=\linewidth]{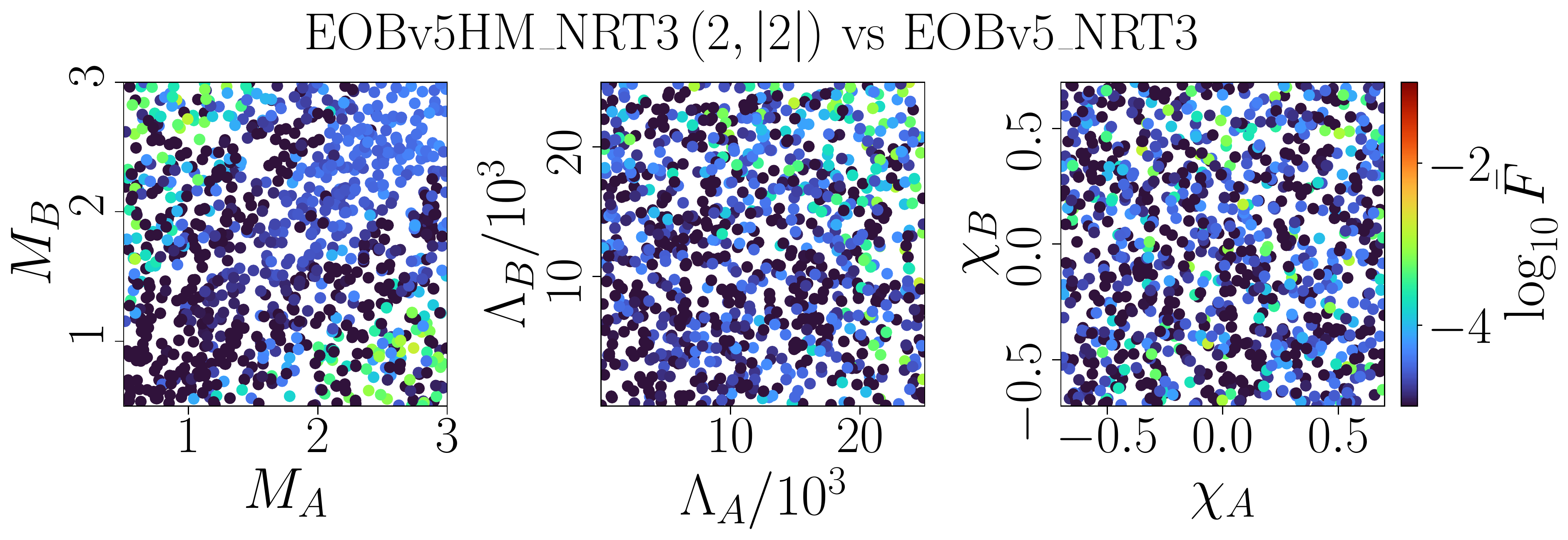}
\caption{Mismatch comparisons of the HM waveform models against their $(2,|2|)$-mode counterparts. Each subfigure contains three two-dimensional, density scatter plots of the masses $M_{A,B}$, tidal deformabilities $\Lambda_{A,B}$, and aligned spin components $\chi_{A,B}$, where the log-mismatch $\log_{10}\bar{F}$ is represented by the color bar, with limits set to $\log_{10}\bar{F} \in [-5, -1]$.}
\label{fig: mm_vs22modes}
\end{figure}
To evaluate the HM models accuracy, we compute their mismatch with respect to the same NR waveforms employed in the TD comparison. Moreover, in order to compare the different HM models in the FD, we calculate their mismatches against other tidal models over a chosen parameter space. The mismatch between two complex waveforms ${h}_1$ and ${h}_2$, is defined as
\begin{equation}\label{eq: mismatch}
\bar{F} = 1 -\!\!\max_{\phi_0, t_c, \psi_p}\!\frac{(h_1(\phi_c, t_c, \psi_p)|h_2)}{\sqrt{(h_1|h_1)(h_2|h_2)}},
\end{equation}
where the overlap
\begin{equation}\label{eq: overlap}
    (h_1|h_2) = 4 {\rm Re} \int_{f_\text{min}}^{f_{\text{max}}}\frac{\tilde{h}_1^{*}(f)\tilde{h}_2(f)}{S_n(f)}df,
\end{equation}
is computed between the chosen minimum and maximum frequencies, $f_{\rm min}$ and $f_{\rm max}$, respectively, with $\tilde{h}(f)$ being the Fourier transform of $h(t)$, and $\tilde{h}^*(f)$ its complex conjugate. The maximization of the overlap for some reference phase $\phi_0$, coalescence time $t_c$ and polarization angle $\psi_p$  
ensures the alignment between the two waveforms. Throughout this section, we assume a flat spectral density, $S_n(f) = 1.0$, so that the computation is detector-agnostic.

\textbf{Mismatches against NR simulations.} First, we compute mismatches between the NR waveforms in Table~\ref{table: bns_td_configs}, and the HM models \xhmnrtthree, \xhmnrttwo, and \eobhmnrtthree. We calculate the mismatches over a grid of $\iota \in \{0, \pi/3, 0.99\pi/2\}$ (from a face-on, to an almost edged-on system), $\psi_p\in \{0, \pi/2, \pi\}$ and $\phi_0\in \{0, \pi/2\}$ and take the average over the 18 different combinations of these parameters. The mismatches are computed for a frequency range $[f_0^{\rm NR}, f_{\rm mrg}]$, where $f_0^{\rm NR}$ is the minimum frequency of the NR simulation. The results are shown in Fig.~\ref{fig: NRmismatches}, where the markers in the plot indicate the mean mismatches between the model and the NR waveform, while the upper and lower whiskers denote the maximum and minimum mismatches, respectively. We compare the NR waveforms with all their available modes with both the HM and $(2,|2|)$-only models. In general, HM models yield lower mean mismatches than their $(2,|2|)$-mode counterparts, indicating that HM models are indeed better suited to describe NR simulations with contributions from multiple modes. 

Regarding the comparison between the different HM models, also in this case, when considering waveforms with multiple modes, we compute the mismatches for the aforementioned combinations of $\iota$, $\phi_0$ and $\psi_p$, and take the average mismatch. We set $f_{\rm min} = 20 \, \rm Hz$, $f_{\rm max} = 2048 \, \rm Hz$, and the sampling rate $T_s = 2f_{\rm max} = 4096 \,\rm{Hz}$. We compute three different sets of mismatches between the various waveform models: (a) mismatches over all modes, including in the computation of the strain using all available modes from each model, (b) mode-by-mode mismatches between individual common modes, and (c) mismatches between the $(2,|2|)$-mode of the HM waveforms, and their $(2,|2|)$-only approximant counterpart (e.g., $(2,|2|)$-mode of \xhmnrtthree\ versus \xasnrtthree). For mismatches involving mode-by-mode comparisons, instead, we set $\iota = \pi/6$, $\phi_0 = 0$, and $\psi_p = 0$. Moreover, for most of these tests, we randomly select 1000 configurations sampled over the parameter space $M_{A,B} = [0.5, 3.0]M_{\odot}$, $\Lambda_{A,B} = [0, 25000]$, and $|\chi_{A,B}| \le 0.7$. The same data samples were used for the mismatch computations to ensure a fair comparison between the models. For the comparison with \texttt{SEOBNRv5THM}, we use a narrower parameter space $M_{A,B} = [1.0, 3.0]M_{\odot}$, $\Lambda_{A,B} = [0, 5000]$, and $|\chi_{A,B}| \le 0.5$, to consider only the regions where this model guarantees a reliable waveform generation.
Table~\ref{table: meanmax_mm} shows the median $\operatorname{med}({\bar{F}})$ and maximum $\max({\bar{F}})$ mismatches over all the source configurations for the different comparisons.

\textbf{Mismatches over all modes.} In this case, we calculate the mismatches between different HM waveform models including all their available modes. We show in Fig.~\ref{fig: mm_allmodes} the mismatches between the different aligned spin \nrtidalvthree\ approximants, and between the \nrtidalvthree\ and \nrtidalvtwo\ tidal waveforms. We note a very good agreement between \xhmnrtthree\ and \eobhmnrtthree\, with mismatches of the order of $\mathcal{O}(10^{-4})$. This is consistent with these models incorporating similar tidal approximants and differing primarily in their BBH baselines and the fact that \eobhmnrtthree\ includes the additional $(4,|3|)$ and $(5,|5|)$ modes. Meanwhile, the mismatches are larger between \eobhmnrtthree\ and \xhmnrttwo\ (${\rm med}(\bar{F}) = 0.0023$) because of the differences not only in the BBH baselines, but also in the tidal models employed, which is confirmed by the relatively large mismatches also between \xhmnrtthree\ and \xhmnrttwo\ (${\rm med} (\bar{F}) = 0.0020$).  
In general, these mismatch values agree with the ones calculated between the different $(2,|2|)$-mode models in Ref.~\cite{Abac:2023ujg}. Comparing \xhmnrtthree\  with \seobnrvfivethm\, we also observe mismatches of $\mathcal{O}(10^{-3})$ because of the differences both in BBH baselines and tidal description. The same holds for \eobhmnrtthree\ vs \seobnrvfivethm, due to a difference in the implementation between the BBH baselines employed in \texttt{SEOBNRv5HM}, on which \seobnrvfivehmrom\ is based, and \texttt{SEOBNRv5THM}. 
Specifically, the mismatches between the two models in the BBH case, i.e., setting $\Lambda=0$, in \seobnrvfivethm\, are already at $\mathcal{O}(10^{-4})$ as tested in Ref.~\cite{Haberland:2025luz}. The comparison of both our HM models with \seobnrvfivethm\ including all the modes are consistent with the findings of Ref.~\cite{Haberland:2025luz} relative to the comparison of the $(2, |2|)$-mode versions of the models.

\textbf{Mode-by-mode mismatches.} 
We also compare each common mode between \xhmnrtthree\ and \eobhmnrtthree, by computing the mismatches mode-by-mode. The results are shown in Fig.~\ref{fig: mode_by_mode_mm} for the different modes. We note that the mismatch values between the $(2,|2|)$-modes of \xhmnrtthree\ and \eobhmnrtthree\ and their distribution over the parameter space are consistent with the mismatches between \xasnrtthree\ and \eobnrtthree\ in Ref.~\cite{Abac:2023ujg}. From Table~\ref{table: meanmax_mm}, we see that in general the mismatches between HMs are roughly an order of magnitude larger than the $(2,|2|)$-mode only ones. This is attributed to the differences in the BBH baselines, since the results here are overall consistent with the mismatch values between the BBH baselines found in Ref.~\cite{Garcia-Quiros:2020qpx}. We also further note that large values of tidal deformabilities and spins yield larger values of the mismatches as expected.

\textbf{Mismatches between $(2,|2|)$-modes.}
Lastly, we compare the HM \nrtidalvthree\ models with their $(2,|2|)$-mode counterparts. We compare \xhmnrtthree\ against \xasnrtthree\ as well as \eobhmnrtthree\ and \eobnrtthree. The results for these comparisons are shown in Fig.~\ref{fig: mm_vs22modes}, where we find that the mismatches are relatively large for large mass ratios and tidal deformabilities, but remain overall small and consistent with the ones observed between the different tidal models altogether, c.f., Fig.~\ref{fig: mm_allmodes}. These larger mismatches are due to the differences in the tapering between the HM models and the purely $(2,|2|)$-mode models, which are discussed in Sec.~\ref{subsection: Post-Merger}\footnote{We confirm this by performing additional tests for the HM models versus the $(2,|2|)$ for which their tapering are identical, and find mismatches of the order of machine precision, i.e. $\mathcal{O}(10^{-16})-\mathcal{O}(10^{-15})$.} .
\begin{table}[t!]
\caption{\label{table: meanmax_mm} 
Median ($\operatorname{med}$) and maxium ($\max$) mismatch between waveform models for several mode configurations. `$-$' means that the mismatch was not computed.}
\centering
\renewcommand{\arraystretch}{1}
\setlength{\tabcolsep}{4pt}
\small
\begin{tabular}{l|c|c|c}
\hline
\hline
\multicolumn{4}{c}{\textbf{All Modes}} \\
\hline
& $\rule{0pt}{10pt}\bar{F}$ & \xhmnrtthree & \eobhmnrtthree\\
\hline
\xhmnrtthree 
    & med & -- & 0.00042 \\
    & max & -- & 0.0053 \\
\eobhmnrtthree 
    & med & 0.00043 & -- \\
    & max & 0.0053 & -- \\
\xhmnrttwo 
    & med & 0.0020 & 0.0023 \\
    & max & 0.016 & 0.015 \\
\texttt{SEOBNRv5THM} 
    & med & 0.0053 & 0.0058 \\
    & max & 0.035 & 0.037 \\
\hline
\hline
\multicolumn{4}{c}{\textbf{Mode-by-Mode: \xhmnrtthree\ vs \eobhmnrtthree}} \\
\hline
Mode & Metric & \multicolumn{2}{c}{Mismatch} \\
\hline
$(2,|2|)$ & med & \multicolumn{2}{c}{0.00016} \\
        & max & \multicolumn{2}{c}{0.0030} \\
$(2,|1|)$ & med & \multicolumn{2}{c}{0.0058} \\
        & max & \multicolumn{2}{c}{0.42} \\
$(3,|2|)$ & med & \multicolumn{2}{c}{0.0015} \\
        & max & \multicolumn{2}{c}{0.038} \\
$(3,|3|)$ & med & \multicolumn{2}{c}{0.0043} \\
        & max & \multicolumn{2}{c}{0.53} \\
$(4,|4|)$ & med & \multicolumn{2}{c}{0.0035} \\
        & max & \multicolumn{2}{c}{0.034} \\
\hline
\hline
\multicolumn{4}{c}{\textbf{$(2,|2|)$-Mode Models}} \\
\hline
& & \xasnrtthree & \eobnrtthree \\
\hline
\xhmnrtthree 
    & med & 0.000016 & -- \\
    & max & 0.0017 & -- \\
\eobhmnrtthree 
    & med & -- & 0.000016 \\
    & max & -- & 0.0017 \\
\hline
\end{tabular}
\end{table}

\section{\label{section: Parameter estimation} Parameter Estimation}
\begin{table}[t]
\centering
\caption{Source parameter injection values for the aligned-spin comparable-mass (\equalas) and high-mass ratio (\unequalas) systems, as well as the precessing \unequalprec\ system. In the system labels, the value after \texttt{M} denoted the total mass of the system ($3.4M_{\odot}$ or $3.5M_{\odot}$), while the value that follows \texttt{q} denotes the mass ratio ($q = 1.05$ or $q = 2.27$).}
\label{table: inj-rec-sim-mass}
\begin{tabular}{l |c|c|c}
\hline
Parameter       &  \equalas & \unequalas & \unequalprec \\
\hline
\hline
$\mathcal{M}_c$ [$M_{\odot}$] & 1.4875 & 1.3772 & 1.3772  \\
$1/q$           &0.95   & 0.44 & 0.44\\
$\Lambda_A$     &373    & 60 & 60\\
$\Lambda_B$     &483    & 3737 & 3737\\
$\chi_A$        &0.02   & 0.08 & - \\
$\chi_B$        &0.01   & 0.16 & - \\
$\theta_{JN}$   & 0.879 & 1.0995 & 1.0995\\
$D_L$ [Mpc]     & 60    & 71 & 71\\
$\delta$        & -1.22 & -0.88 & -0.88\\
$\alpha$        & 1.67  & 2.39 & 2.39\\
$\psi_p$        & 2.70  & 1.45 & 1.45\\
$\phi_c$        & 3.69  & 2.76 & 2.76\\ 
Duration [s]    & 128   & 256 & 256\\
$a_A$ & - & - &0.08\\
$a_B$ & - & - &0.16\\
$\zeta_A$ & - & - &1.36\\
$\zeta_B$ & - & - &2.08\\
$\phi_{AB}$ & - & - &5.919\\
$\phi_{JL}$ & - & - &3.383\\
\hline
\end{tabular}
\end{table}

\begin{table}[t!]
\caption{\label{table: prior} Prior distributions for the different source parameters of the \equalas\ [\unequalas] systems. Most of the parameters are represented by the uniform distribution $\mathcal{U}$. In this table, $\mathcal{U}^{\rm AS}$ corresponds to an aligned-spin prior with a uniform distribution in magnitude, while $\mathcal{U}^{\rm SF}$ is uniform in comoving volume and source frame time~\cite{Ashton:2018jfp}. Additional precessing settings for the \unequalprec\ system are also added in the table.}
\begin{tabular}{l|l} 
\hline
Parameter & Prior Distribution\\
\hline
$\mathcal{M}_c$ [$M_{\odot}$]  & $\mathcal{U} (1.4, 1.6)$ [$\mathcal{U} (1.3, 1.5)$]\\
$ 1/q $       & $\mathcal{U}(0.05, 1.0)$ \\
$\Lambda_A$   & $\mathcal{U}(0, 5000)$ \\
$\Lambda_B$   & $\mathcal{U}(0, 5000)$ \\
$\chi_A$      & $\mathcal{U}^{\rm AS}(0, 0.05)$ [$\mathcal{U}^{\rm AS}(0, 0.4)$]\\
$\chi_B$      & $\mathcal{U}^{\rm AS}(0, 0.05)$ [$\mathcal{U}^{\rm AS}(0, 0.4)$]\\
$D_L$ [Mpc]   & $\mathcal{U}^{\rm SF}(1, 500)$\\
$\delta$      & $\rm Cosine$\\
$\alpha$      & $\mathcal{U}(0, 2\pi)$ (periodic)\\
$\theta_{JN}$ & $\mathcal{U}(0, \pi)$\\
$\psi_p$      & $\mathcal{U}(0, \pi)$ (periodic)\\
$\phi_c$      & $\mathcal{U}(0, 2\pi)$ (periodic)\\
\hline
\multicolumn{2}{c}{Precessing Settings}\\
\hline
$a_A$  & $\mathcal{U}(0, 0.4)$\\
$a_B$  & $\mathcal{U}(0, 0.4)$\\
$\zeta_A$  & ${\rm Sine}$\\
$\zeta_B$  & ${\rm Sine}$\\
$\phi_{AB}$  & $\mathcal{U}(0, 2\pi)$ (periodic)\\
$\phi_{JL}$  & $\mathcal{U}(0, 2\pi)$ (periodic)\\
\hline
\end{tabular}
\end{table}
Finally, we employ our BNS HM models to perform PE analyses for BNS systems with various source configurations. We use \texttt{bilby}~\cite{Ashton:2018jfp} with the sampler \texttt{dynesty}~\cite{Speagle:2019ivv} and 1000 live points. Given some source parameters $\vec{\mathbf{\theta}}$, data $\mathbf{d}$, and hypothesis $\Omega$, Bayes' theorem reads
\begin{equation}
    p(\vec{\mathbf{\theta}} | \mathbf{d}, {\Omega}) = \frac{\mathcal{L}(\mathbf{d}|\vec{\mathbf{\theta}}, {\Omega})\pi(\vec{\mathbf{\theta}}|{\Omega})}{E(\mathbf{d}|\Omega)},
\end{equation}
where $p(\vec{\mathbf{\theta}} | \mathbf{d}, {\Omega})$ is the posterior probability distribution of $\vec{\theta}$ given $\mathbf{d}$ and $\Omega$, $\pi(\vec{\mathbf{\theta}}|{\Omega})$ is the prior probability distribution of the parameters, $E(\mathbf{d}|\Omega)$ is the evidence, which, in this context, serves as a normalization constant to $p(\vec{\mathbf{\theta}} | \mathbf{d}, {\Omega})$, and $\mathcal{L}(\mathbf{d}|\vec{\mathbf{\theta}}, {\Omega})$ is the likelihood of obtaining $\mathbf{d}$ given $\vec{\mathbf{\theta}}$ under the hypothesis $\Omega$~\cite{Ashton:2018jfp, Romero-Shaw:2020owr, Ashton:2021cub}. To speed up the likelihood computation in the analysis, we employ the adaptive frequency resolution method, i.e., multibanding. This method takes advantage of the chirping behavior of the GW signal, dividing its frequency range into multiple bands so that the time-to-merger of each band is smaller than that of the previous ones, and adapting the frequency-grid step accordingly~\cite{Morisaki:2021ngj}. 

We perform parameter estimation both on simulated signals, with injections in zero noise and recovery with the LVK-design sensitivity~\cite{LVK-T2000012-v2}, and on the real event GW170817; marginalizing the likelihood evaluations over the luminosity distance~\cite{Ashton:2018jfp, Singer:2015ema}   .
We consider different types of systems: %
(i) a comparable-mass, aligned-spin system with small spins (labeled \equalas, Sec.~\ref{subsection: injection-recovery with comparable-mass-ratio system}); %
(ii) a high-mass-ratio, aligned-spin system with moderate spins (\unequalas, Sec.~\ref{subsection: injection-recovery with high-mass-ratio system}); %
(iii) a precessing version of the high-mass ratio system (\unequalprec, Sec.~\ref{subsection: injection-recovery with precessing system}); and %
(iv) recovery of GW170817 data (Sec.~\ref{subsection: pe-GW170817}).

For simulated signals, injection and recovery are performed with the same waveform model. Meanwhile, for the precessing case, the injected signal is generated with \xphmnrtthree\, and we recover this signal with \xphmnrtthree\ as well as other models.

In Table~\ref{table: inj-rec-sim-mass}, we provide the injected parameters values for each system. Tidal deformabilities were computed for the systems' masses for a specific EOS from the set in Ref.~\cite{Koehn:2024set}, chosen such that lower masses yield very large tidal deformabilities, which should stress-test the models and the PE pipelines. 
The prior distributions for the various source parameters for the different systems are listed in Table~\ref{table: prior}. 

\subsection{Injection study for an aligned-spin, comparable-mass system}
\label{subsection: injection-recovery with comparable-mass-ratio system}
\begin{figure*}
\centering
\includegraphics[width=\linewidth]{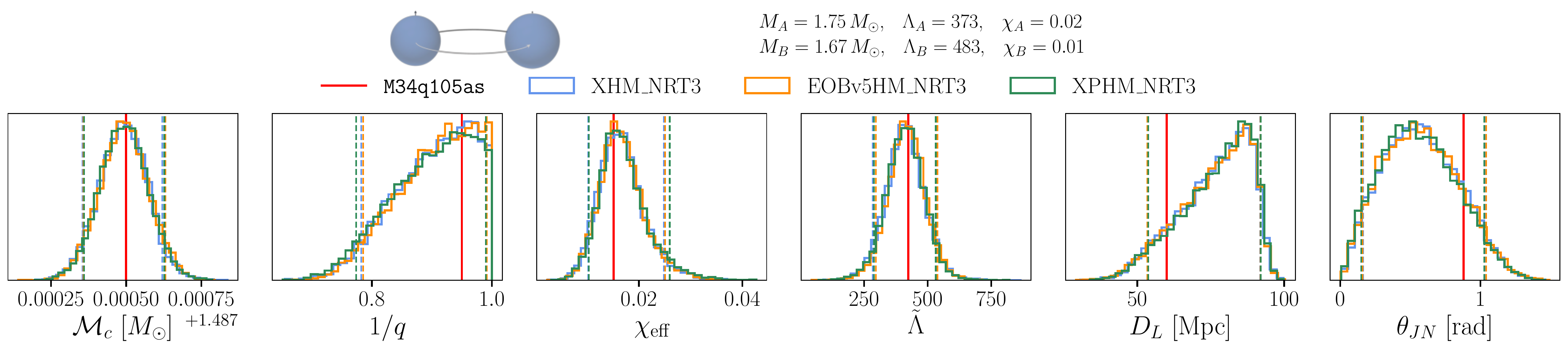}
\caption{Marginalized 1D posterior probability distributions for chirp mass $\mathcal{M}_c$, binary tidal deformability $\tilde{\Lambda}$, effective spin $\chi_{\rm eff}$, luminosity distance $D_L$, and inclination angle $\theta_{JN}$ for the aligned-spin \equalas\ system, injected and recovered with \xhmnrtthree\ (blue), \eobhmnrtthree\ (orange) and \xphmnrtthree\ (green). Vertical dashed lines in the 1D plots indicate 90\% confidence interval. The injected values are indicated by red lines. We also put above the 1D-plots a sketch of the system, set to-scale, along with its source properties.}
\label{fig: corner-A-128}
\end{figure*}
The results for the injection-recovery for the \equalas\ system are shown in Fig.~\ref{fig: corner-A-128} as 1D marginalized posterior distributions of $\mathcal{M}_c$, $1/q$, the binary mass-weighted tidal deformability $\tilde{\Lambda}$, the effective spin $\chi_{\rm eff}$, and luminosity distance $D_L$. The binary tidal deformability $\tilde{\Lambda}$ is defined in terms of the individual masses and tidal deformabilities of the stars~\cite{Dietrich:2020efo}:
\begin{equation}
    \tilde{\Lambda} = \frac{16}{13}\frac{(M_A + 12M_B)M_A^4\Lambda_A + (M_B + 12M_A)M_B^4\Lambda_B}{(M_A + M_B)^5},
\end{equation}
while the effective spin $\chi_{\rm eff}$ is defined as
\begin{equation}
    \chi_{\rm eff} = \frac{M_A \chi_A + M_B\chi_B}{M}.
\end{equation}
We observe from Fig.~\ref{fig: corner-A-128} that the injected values of the source parameters are recovered well by the three models, with the injection, indicated by red vertical lines, lying inside the 90\% confidence interval of the posterior distributions. This is also supported by the nearly equal mass of the injected system, which reduces the relative contribution of the odd $m$ modes to the overall waveform. There is a noticeable bias in the recovered luminosity distance for both models toward larger values, but this is consistent with the results found in Ref.~\cite{Abac:2023ujg} and Ref.~\cite{Colleoni:2025aoh}.

\begin{figure}
\centering
\includegraphics[width=\linewidth]{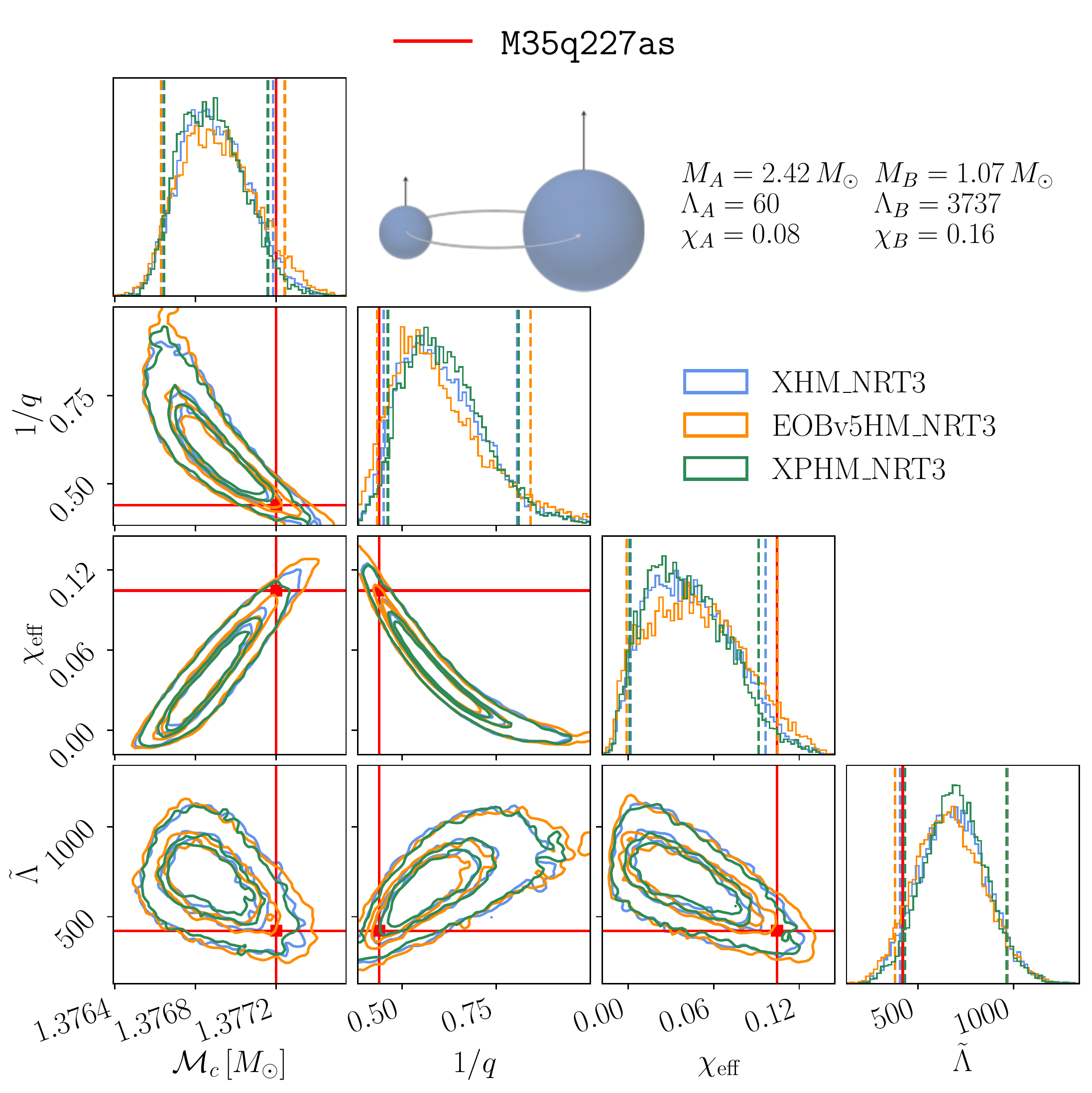}
\caption{The marginalized 1D and 2D posterior probability distributions for selected parameters of the \unequalas\ system, injected and recovered with \xhmnrtthree\ (blue), \eobhmnrtthree\ (orange) and \xphmnrtthree\ (green). The parameters shown here are the chirp mass $M_c$, binary tidal deformability $\tilde{\Lambda}$, and effective spin $\chi_{\rm eff}$. The 68\% and 90\% confidence intervals are indicated by contours for the 2D posterior plots, while vertical dashed lines in the 1D plots indicate 90\% confidence interval. For \xphmnrtthree, we use additional precessing spin parameters in the priors. We also overlay a sketch of the system, set to-scale, along with its source properties.}
\label{fig: corner-B-256s}
\end{figure}
\begin{figure}
\centering
\includegraphics[width=\linewidth]{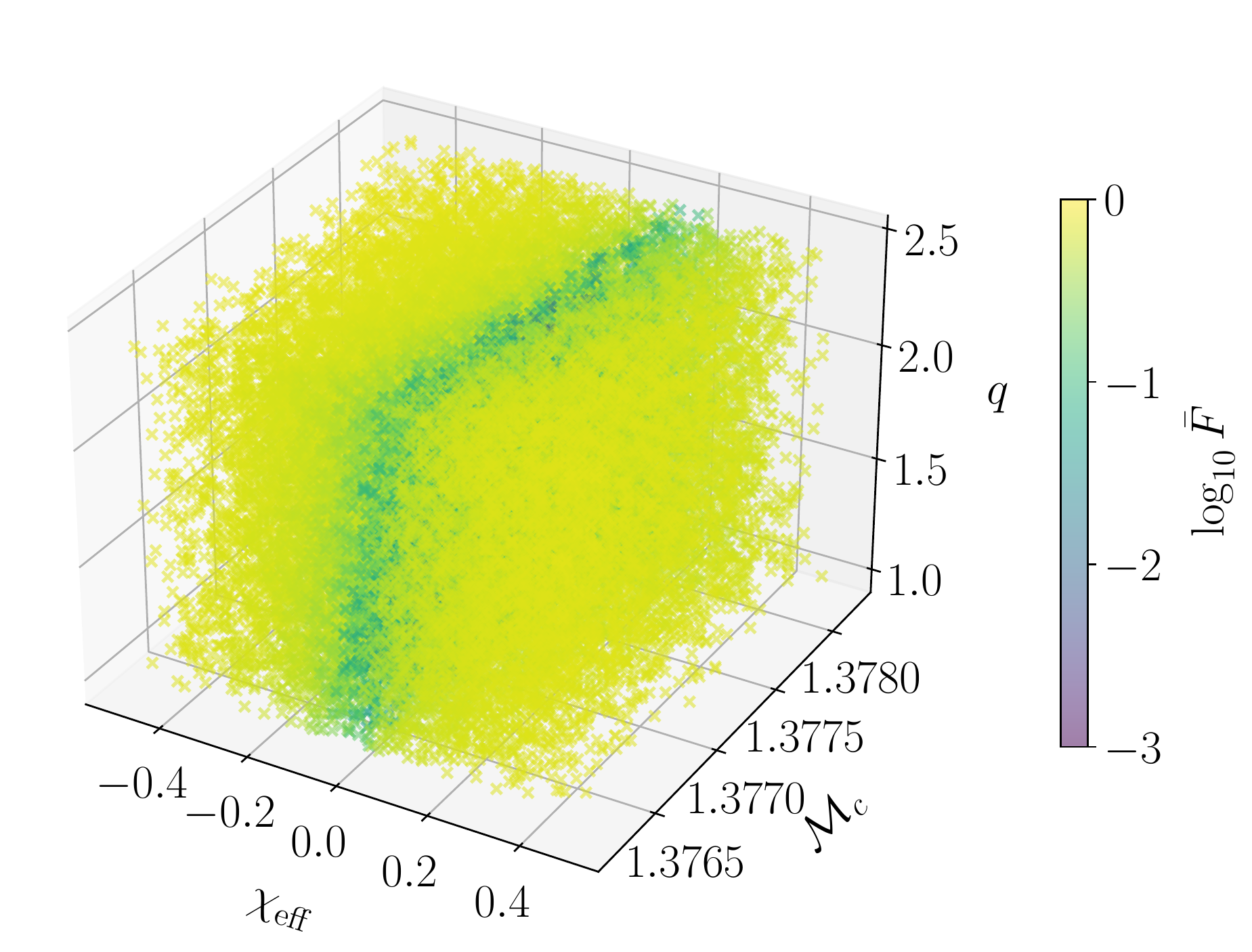}
\caption{Mismatches between \xhmnrtthree---sampled across variations in chirp mass $\mathcal{M}_c$, effective spin $\chi_{\rm eff}$, and mass ratio $q$---and the \unequalas\ system.}
\label{fig: 3d mismatch}
\end{figure}
\begin{figure*}[t]
\centering
\includegraphics[width=0.8\linewidth]{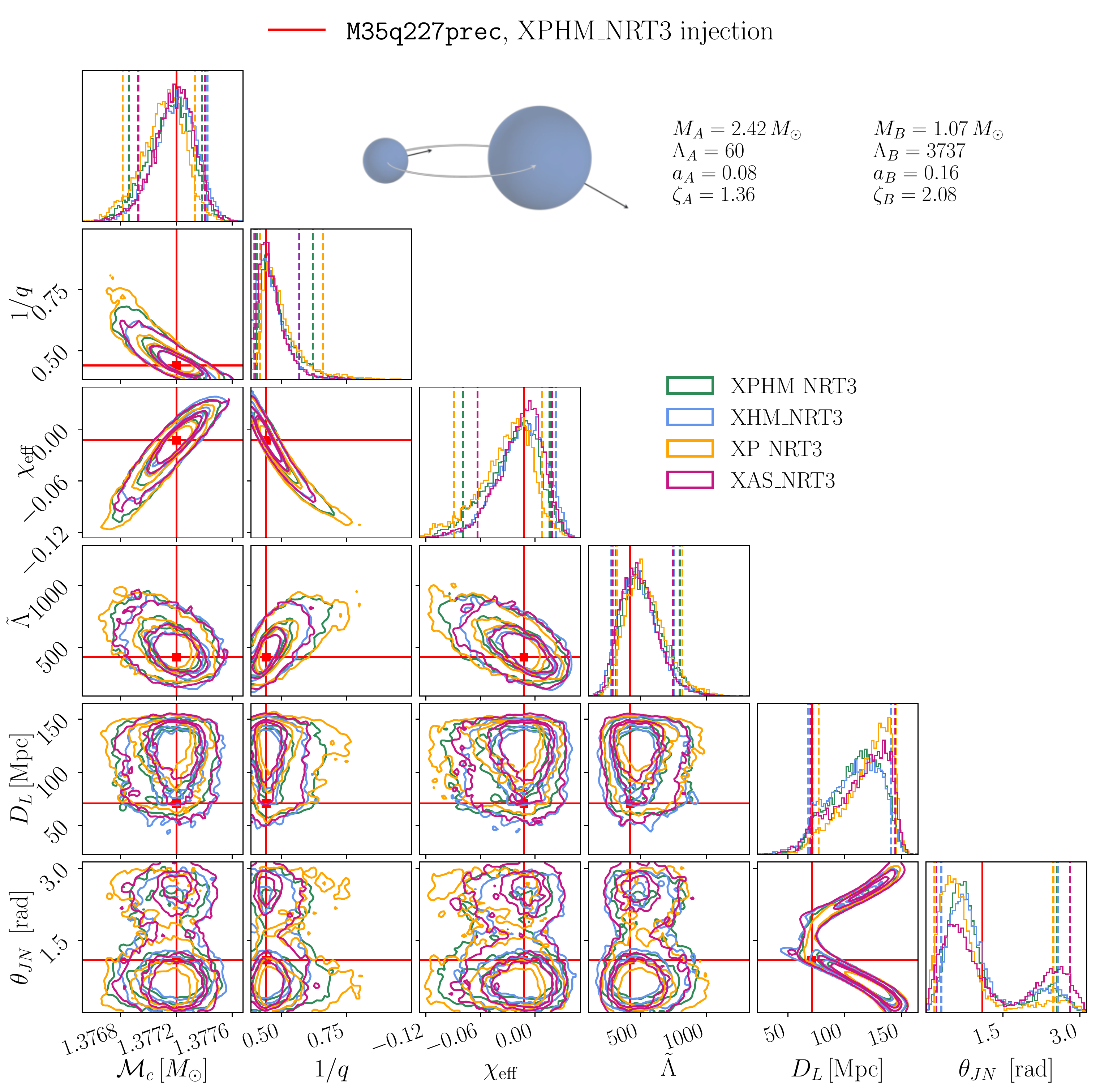}
\caption{The marginalized 1D and 2D posterior probability distributions for selected parameters of the \unequalprec\ system, injected using \xphmnrtthree\ and recovered with \xphmnrtthree, \xpnrtthree, and \xasnrtthree. 
The 68\% and 90\% confidence intervals are indicated by contours for the 2D posterior plots, while vertical dashed lines in the 1D plots indicate 90\% confidence interval. We also overlay a sketch of the system, set to-scale, along with its source properties.}
\label{fig: corner-B-256s-prec}
\end{figure*}

\subsection{Injection study for an aligned-spin, high-mass-ratio system}
\label{subsection: injection-recovery with high-mass-ratio system}
We show the corner plot results for the \unequalas\ system in Fig.~\ref{fig: corner-B-256s}. Unlike the case of \equalas, we find that the recovered posteriors show only marginal agreement with the injected values.

However, this behavior is not due to the addition of HMs to the models. Similar biases are also found with tidal models containing only the $(2,|2|)$-mode (i.e., \xasnrtthree, \eobnrtthree, \xasnrttwo, \imrphenomdnrtidaltwo, and \texttt{SEOBNRv4T\_surrogate}), as shown in Appendix~\ref{subsection: appB}. 
Furthermore, one can also rule out that this problem is intrinsic to tidal models. First, reducing the spins and narrowing the spin priors improves the recovery of the masses and tidal deformabilities; second, also BBH HM baselines demonstrate this degeneracy; and third, as we shall see later, introducing spin precession parameters breaks the degeneracy. We conclude that the reason behind the only marginal parameter recovery of this aligned-spin system is the strong degeneracy between some of the parameters, in particular the masses and spins.

To further explore such degeneracy, we calculate the mismatches between the \xhmnrtthree\ waveform for the \unequalas\ system and for configurations in the parameter space defined by $\mathcal{M}_c \in [1.3762, 1.3782]$  and $|\chi_{A,B}| \le 0.5$, across 31 values of $q$. For each $q$ value, we draw 1000 random configurations of masses and spins, and compute the corresponding $\Lambda_{A,B}$ from the same EOS chosen for PE injections. The choice of the $\mathcal{M}_c$ range reflects the narrow posterior distribution that is typically recovered by PE analyses, see Fig.~\ref{fig: corner-B-256s}. 

The 3D mismatch plot in Fig.~\ref{fig: 3d mismatch} clearly shows the degeneracy between $\mathcal{M}_c$, $\chi_{\rm eff}$, and $q$ across the very narrow $\mathcal{M}_c$ range, with low mismatches found in the $\mathcal{M}_c-q$ plane around the injected $\chi_{\rm eff}$ value. While the degeneracy between $q$ and $\chi_{\rm eff}$, which is also clearly visible in their 2D posterior distribution in Fig.~\ref{fig: corner-B-256s}, is well known~\cite{Colleoni:2025aoh}, Fig.~\ref{fig: 3d mismatch} shows how this extends to a degeneracy between $\mathcal{M}_c$ and $\chi_{\rm eff}$. This explains why, although chirp mass is constrained in a very narrow range, we still see biases in the posterior distribution peak.

\subsection{Injection study for a precessing system}
\label{subsection: injection-recovery with precessing system}
In addition, we inject the precessing \unequalprec\ system using the \xphmnrtthree\ waveform and recover the signal with the same model, \xphmnrtthree\, with the model without precession, i.e., using \xhmnrtthree, with the model without HM content (\xpnrtthree), and, finally, with the model without both precession and HM (\xasnrtthree). 

The corner plots for these runs are shown in Fig.~\ref{fig: corner-B-256s-prec}. As mentioned previously, the introduction of the spin-precessing parameters breaks the degeneracies that were present in the \unequalas\ recovery (Fig.~\ref{fig: corner-B-256s}), and the injected values are generally recovered well by all the models. 
Moreover, the posterior distributions of the precessing models are wider than the aligned-spin ones, due to the introduction of more parameters in the prior to be sampled.
\subsection{GW170817}
\label{subsection: pe-GW170817}
\begin{figure}
\centering
\includegraphics[width=\linewidth]{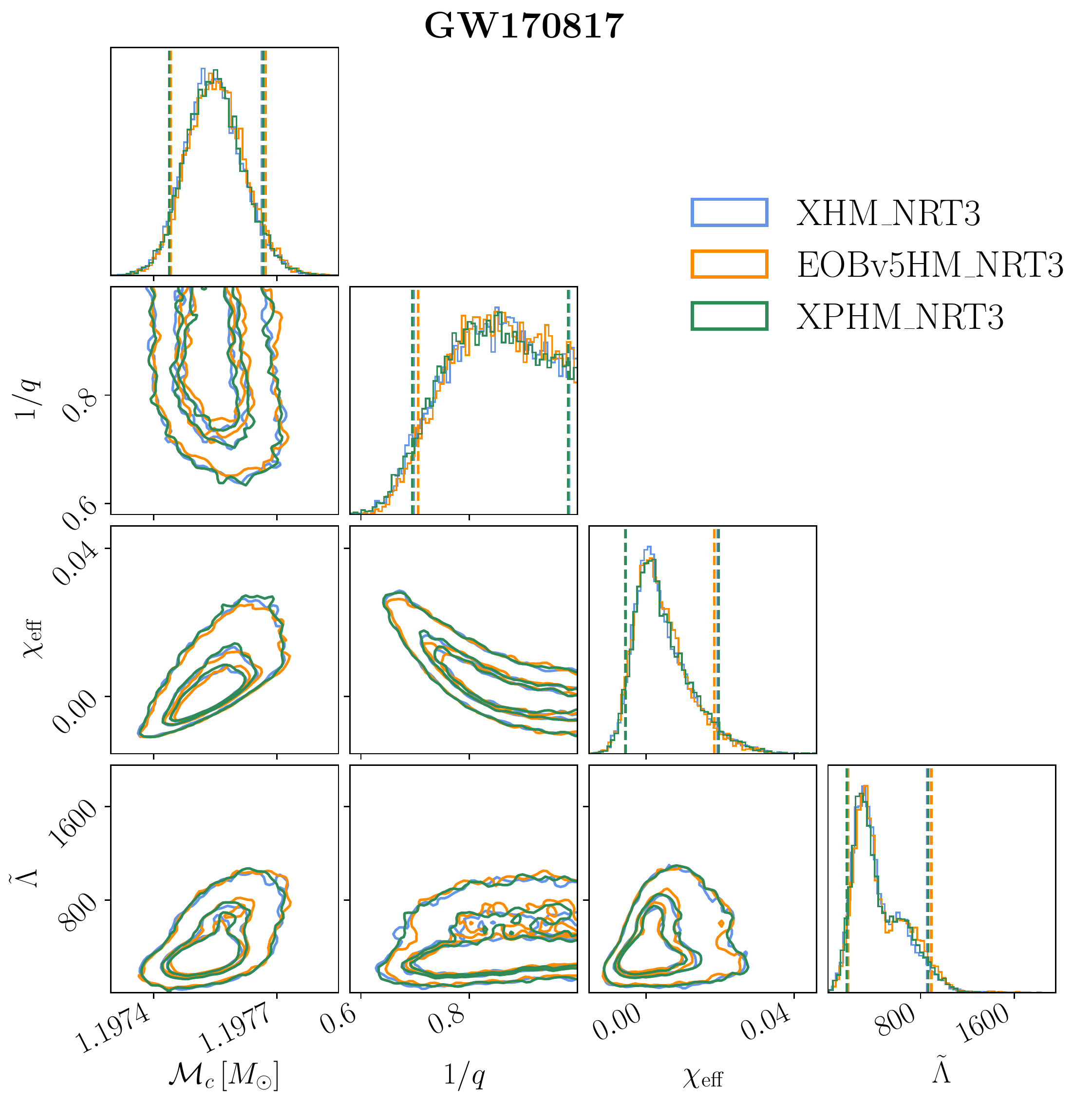}
\caption{The marginalized 1D and 2D posterior probability distributions for selected parameters of GW170817, obtained with \xhmnrtthree\ (blue), \eobhmnrtthree\ (orange), and \xphmnrtthree\ (green). The parameters shown here are the chirp mass $M_c$, effective spin $\chi_{\rm eff}$, and binary tidal deformability $\tilde{\Lambda}$. The 68\% and 90\% confidence intervals are indicated by contours for the 2D posterior plots, while vertical dashed lines in the 1D plots indicate 90\% confidence interval. For \xphmnrtthree, we use additional precessing spin parameters in the priors.}
\label{fig: corner-GW170817}
\end{figure}
Finally, we reanalyze GW170817~\cite{LIGOScientific:2017vwq} using our HM models and compare the results with the analysis performed with the $(2,|2|)$-mode models~\cite{LIGOScientific:2018hze, Abac:2023ujg}. We utilize the same settings as the ones employed for the analysis in the first Gravitational-Wave Transient Catalog (GWTC-1)~\cite{LIGOScientific:2018hze, LIGOScientific:2017vwq}, considering a low-spin prior~\cite{Abac:2023ujg}. The corner plots for this run are shown in Fig.~\ref{fig: corner-GW170817}. We note the consistency between the posterior distributions of the source parameters for the HM waveforms and their corresponding $(2,|2|)$-mode counterparts, c.f., Fig. 16 of Ref.~\cite{Abac:2023ujg}. Such consistency is expected since, as in the case of the \equalas\ system, the HMs do not contribute significantly because the system is nearly symmetric in its component masses. As in the case of the $(2,|2|)$-mode models, we notice a decrease in the secondary peak, and a slightly narrower constraint, for $\tilde{\Lambda}$, compared to the results for GWTC-1, which were obtained with \texttt{IMRPhenomPv2\_NRTidal}~\cite{LIGOScientific:2018hze}.

\section{\label{section: Conclusions and Outlook} Conclusions and Outlook}
In this work, we have extended the \nrtidalvthree\ model to include higher-order mode corrections to the \nrtidalvthree\ phase, exploiting the fact that the higher-order modes phase is approximately proportional to the dominant $(2,|2|)$-mode one. This is the first time that an NR-calibrated, phenomenological tidal model has been extended to include such subdominant corrections. 

The HM version of the \nrtidalvthree\ model was appended to the \imrphenomxhm\ and \seobnrvfivehmrom\ baseline models, which also contain HM extensions. We test the validity of the models in both the time and frequency domains. In the time-domain, we perform a dephasing comparison between different HM models and NR simulations across all their common modes, finding a good agreement. In the frequency-domain, we compute mismatches against NR simulations and between the different models considering various mode configurations. We find that the HM models agree better with the NR simulations than just their $(2,|2|)$-mode counterparts. In general, the \xhmnrtthree\ and \eobhmnrtthree\ models are in good agreement with each other, while for the mode-by-mode comparisons the subdominant modes have a $\mathcal{O}(10^{-1})$ larger mismatch with respect to the $(2,|2|)$-mode comparison.

We performed parameter estimation analyses for different BNS systems using the new models. The injection runs for a comparable-mass system yield very good parameter recovery with the HM models, whereas the injection runs for a particularly high-mass-ratio system result in posterior distributions that only marginally agree with the injected values. Further investigation into this bias, which also occurs for $(2,|2|)$-mode only models, reveals a degeneracy between the chirp mass, mass ratio, and effective spin parameters at a scale comparable to the typical range of chirp masses in the recovered posterior distribution. Finally, we employ the new models to perform PE on the GW event GW170817, and find consistent results with previous analyses.

In the future, this HM version of \nrtidalvthree\ can be employed in the development of NSBH models. The latest NSBH models~\cite{Thompson:2020nei, Matas:2020wab} contain only $(2,|2|)$-mode tidal corrections from \nrtidalvtwo, but the HM content becomes more important for these systems, with highly asymmetric masses. Therefore, NSBH models can be updated using \nrtidalvthree\ with HM corrections and refined by adding tidal amplitude corrections that account for potential tidal disruptions, which are likely to occur for almost-equal-mass systems or aligned-spins systems with large black hole spins.

The newly-developed BNS HM models can be extended in several ways to further increase their accuracy. First, given that the scaling relation of the HM phase with respect to the $(2,|2|)$-phase is an approximation, in principle, one could calibrate the higher-order modes to NR waveforms, with a procedure analogous to the one followed for the $(2,|2|)$-mode in the \nrtidalvthree\ model. This would increase the accuracy of the models, but at the cost of computational efficiency. 

Second, existing $(2,|2|)$-models struggle to describe NR waveforms with relatively large aligned or anti-aligned spin components~\cite{Kuan:2024jnw, Kuan:2025bzu}, hence an update to the \nrtidal\ models would come from calibration to these systems. This could be accomplished through updated prescription for the dynamical tidal effects~\cite{Steinhoff:2021dsn}, which up to this point excludes higher-order spin corrections and, for \nrtidalvthree, only includes linear dynamical tidal corrections for the quadrupolar tidal deformability $\Lambda_2$.

Other physical effects that can be included in the model would be eccentric effects~\cite{Neuweiler:2025lte}, which would modulate the waveform amplitude, as well as reduce the dependence of the model on universal relations. Calibration to more extreme systems, such as those including  subsolar components~\cite{Markin:2023fxx} or those with more exotic EOS~\cite{Giangrandi:2025rko}, would also ensure that future \nrtidal\ models (or BNS models in general) will cover a larger, more reliable, parameter space for the analysis of future detections.

\section*{\label{section: Acknowledgments}Acknowledgments}

This material is based upon work supported by the NSF's LIGO Laboratory which is a major facility fully funded by the National Science Foundation. The authors thank Marcus Haberland, for the assistance he provided in setting up the \texttt{SEOBNRv5THM} waveform interface and for fruitful discussions. The authors also thank Lorenzo Pompili, Alessandra Buonanno, and Jan Steinhoff for helpful discussions. We acknowledge funding from the Daimler and Benz Foundation for the project ``NUMANJI" and from the European Union (ERC, SMArt, 101076369). Views and opinions expressed are those of the authors only and do not necessarily reflect those of the European Union or the European Research Council. Neither the European Union nor the granting authority can be held responsible for them. The parameter estimation runs were performed using the hypatia cluster at the Max Planck Institute for Gravitational Physics. 

\appendix
\section{\label{subsection: appA}Additional TD comparisons with Aligned-Spin NR simulations}
\begin{figure*}
 \centering
\includegraphics[width=0.49\linewidth]{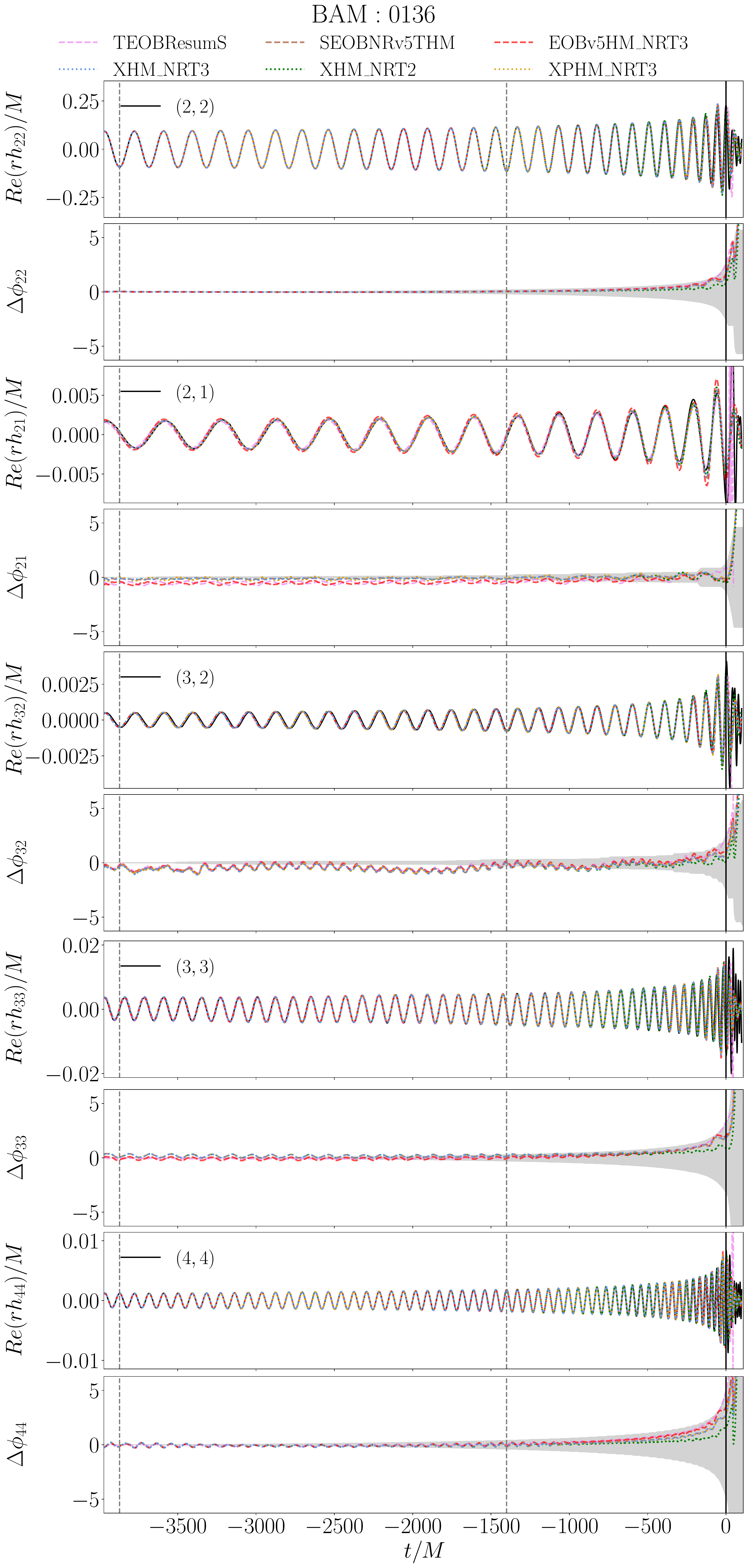}\hfill
\includegraphics[width=0.49\linewidth]{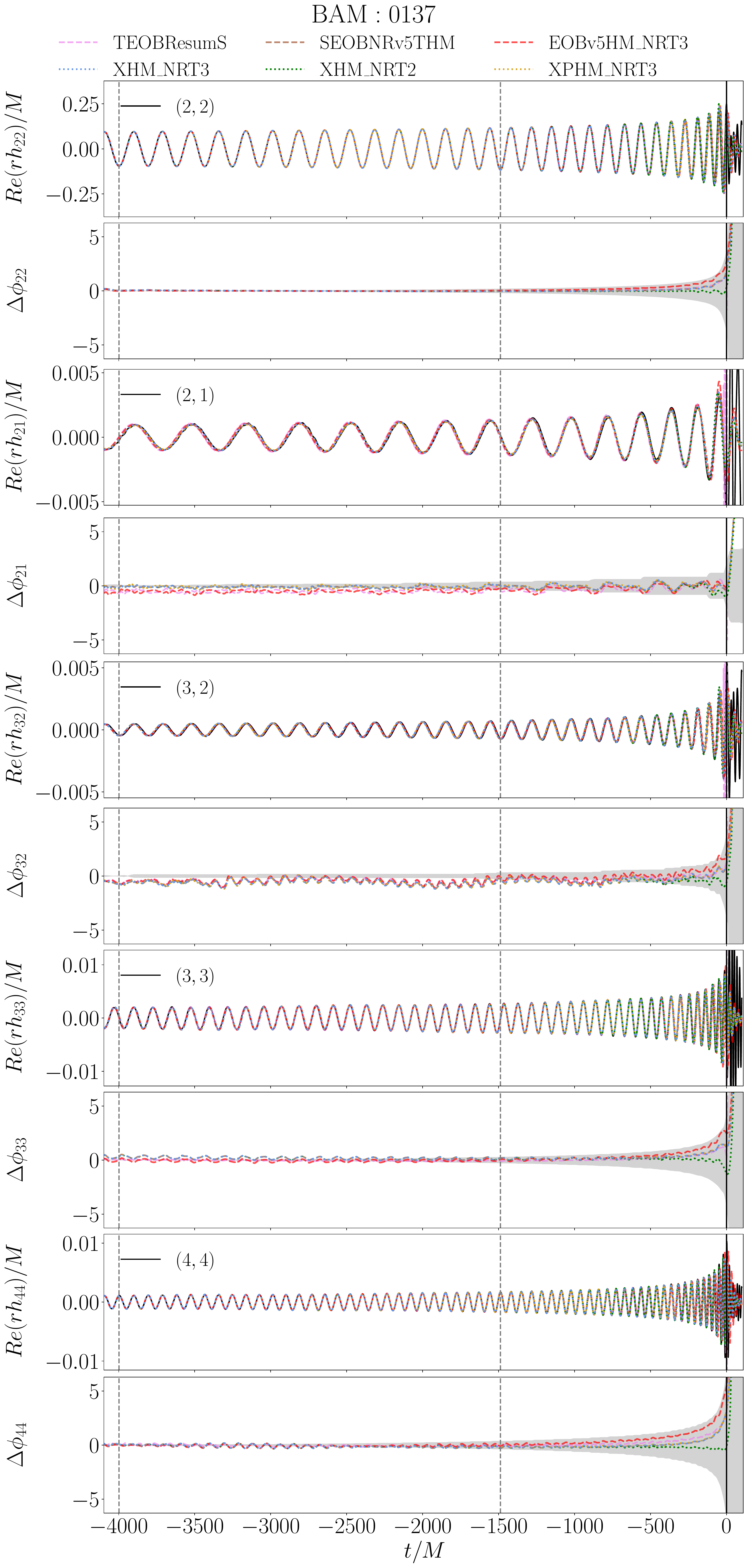}
\caption{TD dephasing comparisons for the BAM:0136 and BAM:0137 waveforms. For each NR waveform, the upper panel shows the real part of the gravitational wave strain as a function of the retarded time, while the bottom panel shows the phase difference between the waveform model and the NR waveform. The gray band in the bottom panel represents the phase difference between the two highest-resolution phases of the simulation. For each comparison, we denote the alignment windows by the dashed gray lines, and merger by the solid black line at $t/M = 0$.}
\label{fig: timedomaincomparisons2}
\end{figure*}

\begin{figure*}
 \centering
\includegraphics[width=0.47\linewidth]{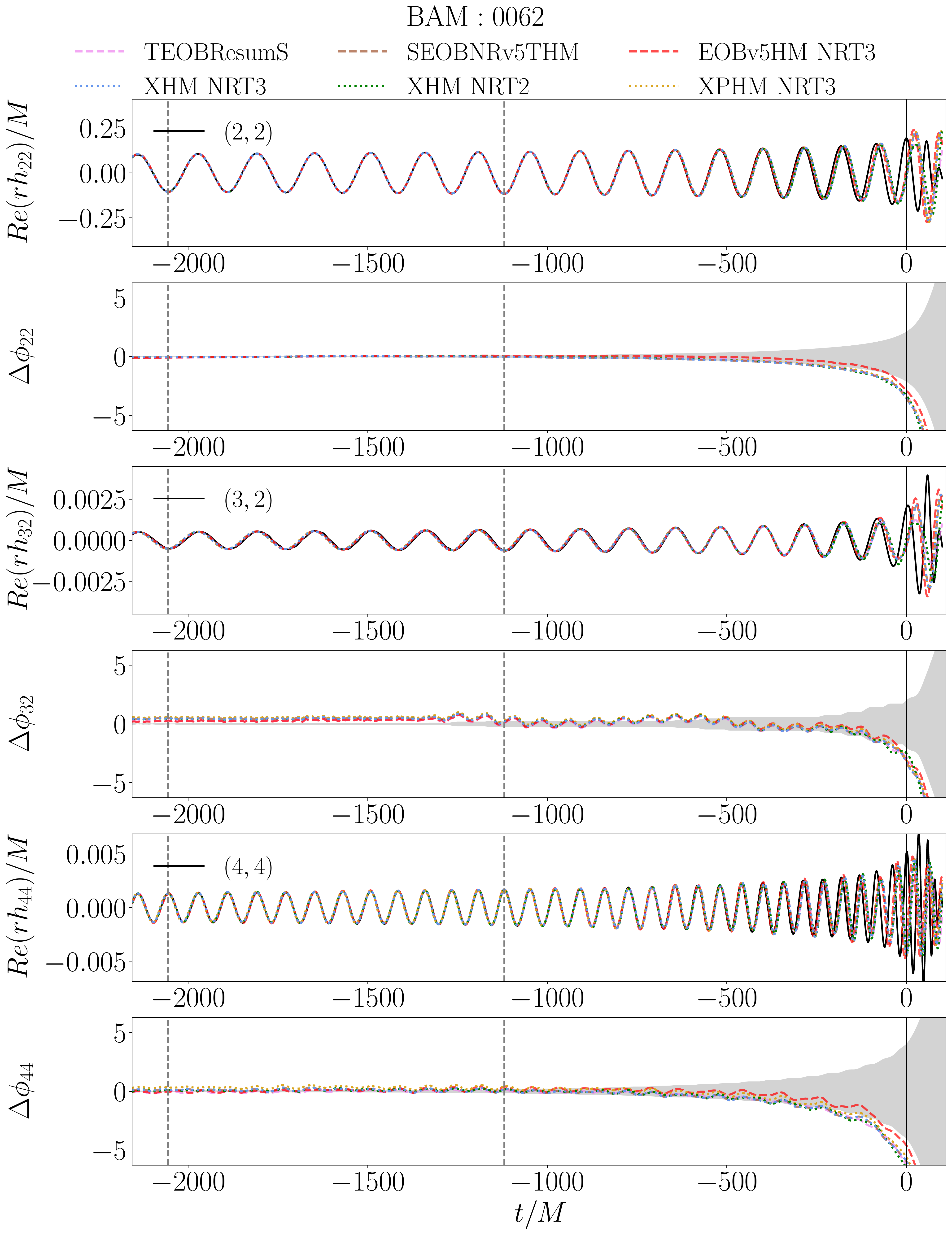}
\includegraphics[width=0.47\linewidth]{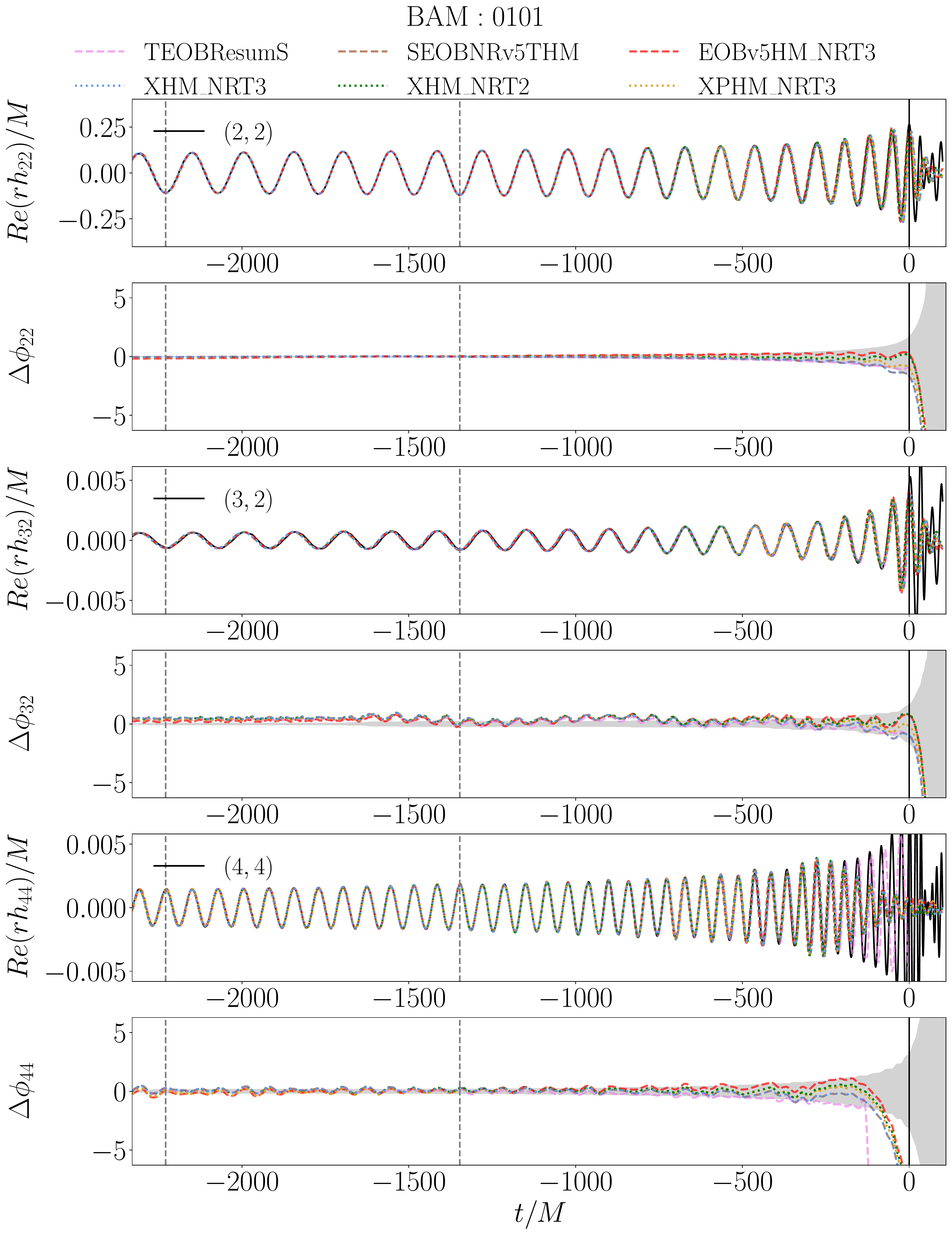}
\caption{TD dephasing comparisons for the BAM:0062 and BAM:0101 waveforms. For each NR waveform, the upper panel shows the real part of the gravitational wave strain as a function of the retarded time, while the bottom panel shows the phase difference between the waveform model and the NR waveform. The gray band in the bottom panel represents the phase difference between the two highest-resolution phases of the simulation. For each comparison, we denote the alignment windows by the dashed gray lines, and merger by the solid black line at $t/M = 0$.}
\label{fig: timedomaincomparisons3}
\end{figure*}

The time domain comparisons for the unequal-mass simulations BAM:0136 and BAM:0137, and equal-mass simulations BAM:0062 and BAM:0101 listed in Table~\ref{table: bns_td_configs} are shown in Fig.~\ref{fig: timedomaincomparisons2} and Fig.~\ref{fig: timedomaincomparisons3}. The Richardson-extrapolated $(2,|2|)$-modes of BAM:0136 and BAM:0137 were also used in the calibration for NRTidalv3, but since the HMs of these simulations do not show a clear convergence, we simply use the highest resolutions for all modes in the comparisons. The same holds for BAM:0062 and BAM:0101. The uncertainty bands in this case represent the phase difference between the two highest resolutions of the simulation. For BAM:0101, we note that the $(4,4)$-mode waveforms from HM BNS models merge earlier than the corresponding NR simulations\footnote{The $m=4$ mode merges at roughly twice the frequency of the $m=2$ mode. This scaling already applies to the HM NRTidalv3 phase. Further adjustments to the taper did not affect the merger time of the $(4, |4|)$-mode for BAM:0101 in Fig.~\ref{fig: timedomaincomparisons3}, (even if they significantly affect the $(4,|4|)$-mode merger for the other simulations). Therefore, we kept the original amplitude taper.}. As an additional note, the \texttt{TEOBResumS-Dalí} waveforms end abruptly at merger by construction, or when the waveform generator returns \texttt{nan} values\footnote{The model is complemented during Bayesian inferences with the BNS frequency domain postmerger model of Ref.~\cite{Breschi:2022xnc}}. For the $(4,|4|)$ mode of BAM:0101 in Fig.~\ref{fig: timedomaincomparisons3}, the phase of the \teobresums\ peaks around $t/M \approx -200$, then abruptly changes slope and becomes \texttt{nan} shortly after. This behavior causes a the sharp deviation in the phase difference relative to the NR waveform. Overall, however, the individual modes of these simulations also agree well with the models.

\section{\label{subsection: appA1}TD comparisons with precessing NR simulations}
\begin{figure*}
 \centering
\includegraphics[width=0.47\linewidth]{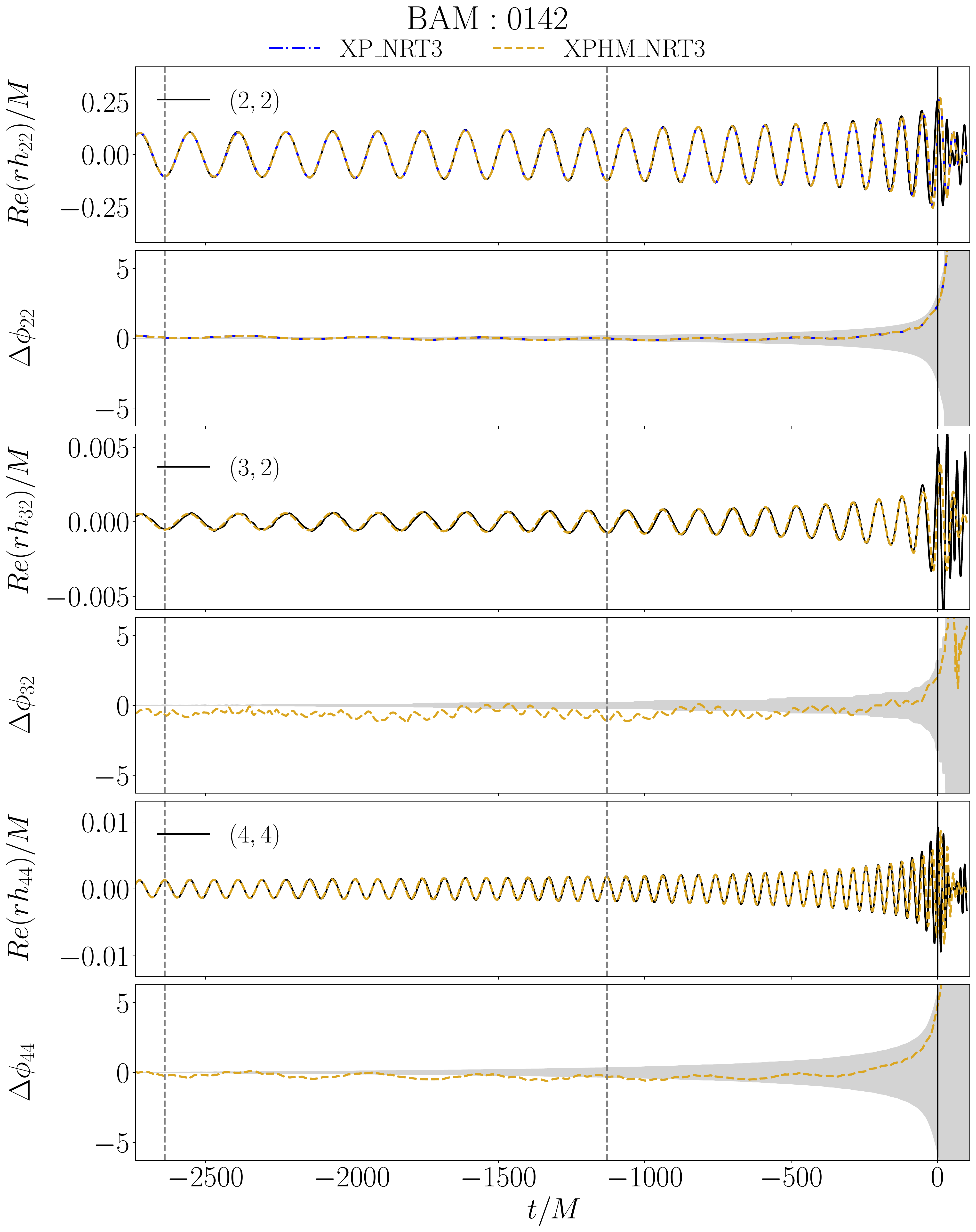}
\includegraphics[width=0.47\linewidth]{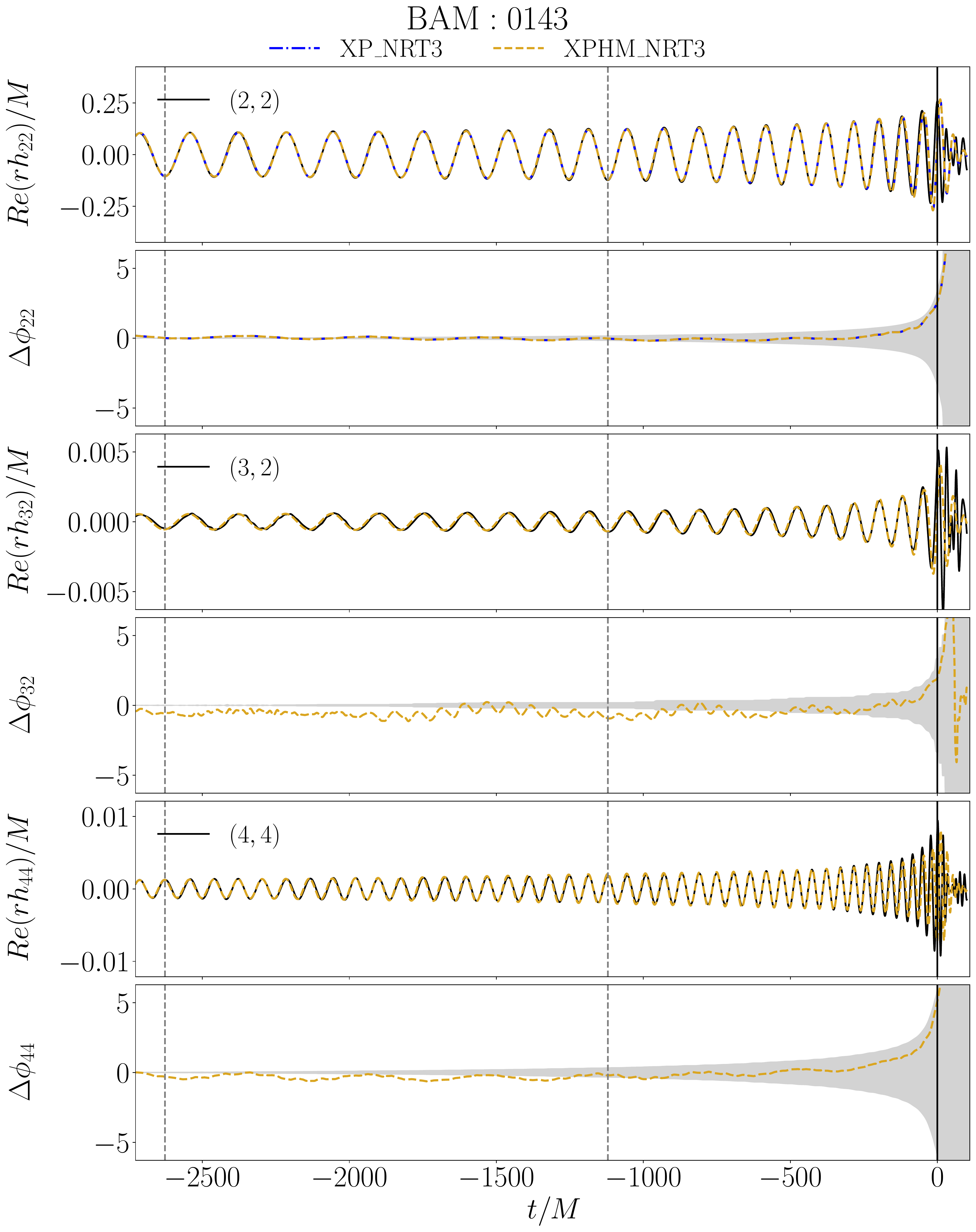}
\caption{TD dephasing comparisons for the precessing BAM:0142 and BAM:0143 waveforms, with models \xphmnrtthree\ and \xpnrtthree. For each NR waveform, the upper panel shows the real part of the gravitational wave strain as a function of the retarded time, while the bottom panel shows the phase difference between the waveform model and the NR waveform. The gray band in the bottom panel represents the phase difference between the two highest-resolution phases of the simulation. For each comparison, we denote the alignment windows by the dashed gray lines, and merger by the solid black line at $t/M = 0$.}
\label{fig: timedomaincomparisons4}
\end{figure*}
In addition to comparisons with the NR simulations in Table~\ref{table: bns_td_configs}, we also compare the precessing models \xphmnrtthree\ and \xpnrtthree\ with the precessing NR simulations BAM:0142 and BAM:0143. Both systems have $M_A=M_B=1.35M_{\odot}$ and $\Lambda_A=\Lambda_B =390$. BAM:0142 has spin components $\vec{\chi}_A = \vec{\chi}_B = [0.068, 0, -0.068]$ while BAM:0143 has $\vec{\chi}_A = \vec{\chi}_B = [-0.068, 0, -0.068]$. We show the results in Fig.~\ref{fig: timedomaincomparisons4}. In general the \xphmnrtthree\ and \xpnrtthree\ models agree well with the NR simulations in the $(2,|2|)$-mode, while \xphmnrtthree\ also agrees well with the HM of the simulations.

\section{\label{subsection: appB}Injection study of \unequalas\ system with $(2,|2|)$-mode models}
\begin{figure}
\centering
\includegraphics[width=\linewidth]{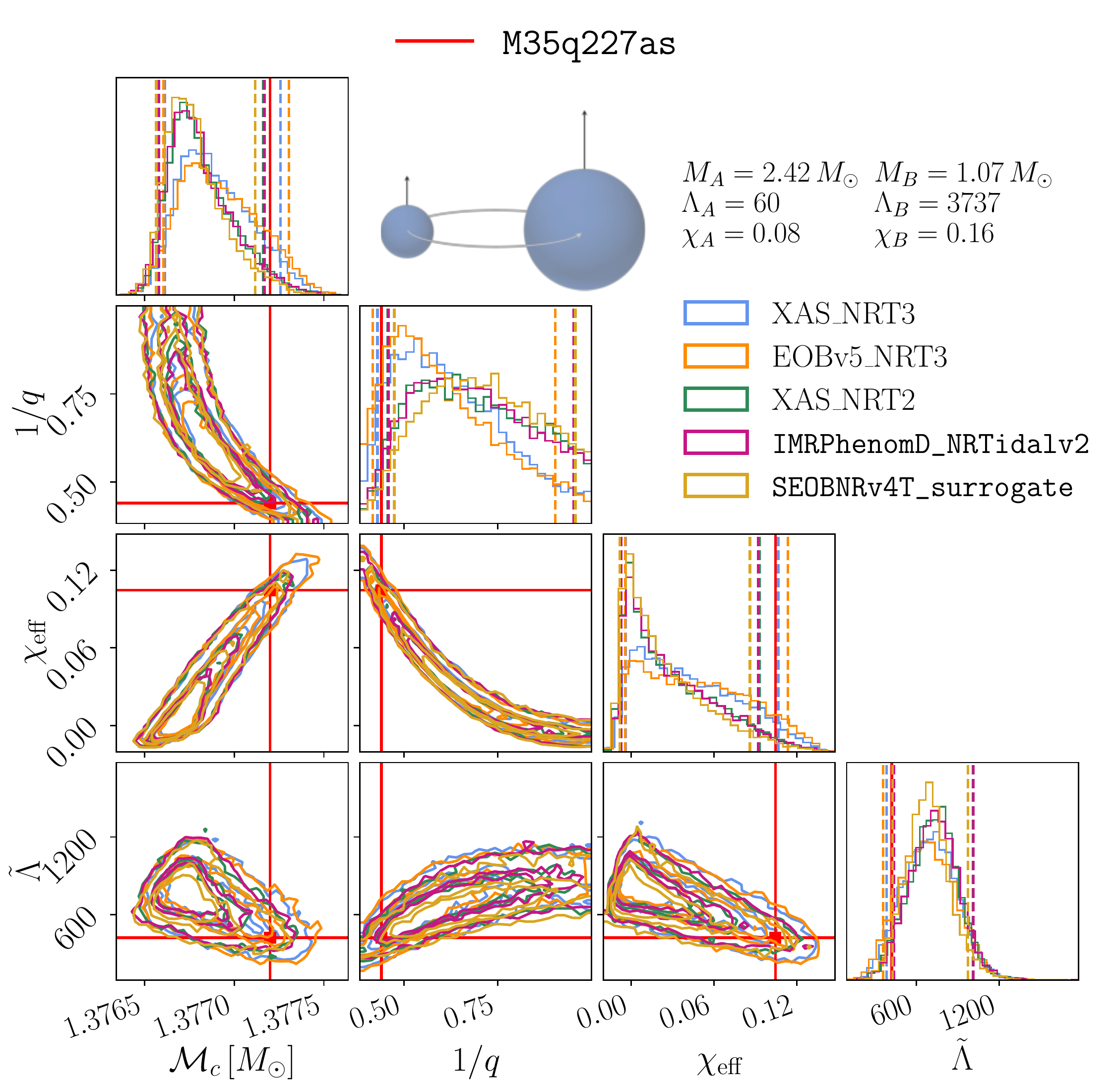}
\caption{The marginalized 1D and 2D posterior probability distributions for selected parameters of the \unequalas\ system, injected and recovered with existing $(2,|2|)$-mode models. In general, the \nrtidalvthree\ models recover the injections better than the \nrtidalvtwo\ models. We note the existence of the degeneracies between the masses and spins that were also present in the aligned-spin HM models, c.f., Fig~\ref{fig: corner-B-256s}.}
\label{fig: corner-B-256s-22}
\end{figure}

To probe that the biases in the recovered parameters for \unequalas\ are not due to the addition of HMs to the waveform models, in Fig.~\ref{fig: corner-B-256s-22} we show the results for the \unequalas\ system injected and recovered with existing $(2,|2|)$-models, similar as in Sec.~\ref{subsection: injection-recovery with high-mass-ratio system}.
We still observe the presence of the degeneracies between the masses and spins for all the models eemployed for recovery, although the \nrtidalvthree\ models in general recover the injected values better than \nrtidalvtwo\ ones. Therefore, we conclude that the degeneracies and the consequent biases in the parameters recovered are not due to the presence of the HM corrections (see Sec.~\ref{subsection: injection-recovery with high-mass-ratio system} for further discussion).

\newpage
\bibliography{References}

\end{document}